\documentclass[12pt]{article}
\input epsf.sty

\begin{document}

\def\CA{{\cal A}}
\def\CB{{\cal B}}
\def\CC{{\cal C}}
\def\CD{{\cal D}}
\def\CE{{\cal E}}
\def\CF{{\cal F}}
\def\CG{{\cal G}}
\def\CH{{\cal H}}
\def\CI{{\cal I}}
\def\CJ{{\cal J}}
\def\CK{{\cal K}}
\def\CL{{\cal L}}
\def\CM{{\cal M}}
\def\CN{{\cal N}}
\def\CO{{\cal O}}
\def\CP{{\cal P}}
\def\CQ{{\cal Q}}
\def\CR{{\cal R}}
\def\CS{{\cal S}}
\def\CT{{\cal T}}
\def\CU{{\cal U}}
\def\CV{{\cal V}}
\def\CW{{\cal W}}
\def\CX{{\cal X}}
\def\CY{{\cal Y}}
\def\CZ{{\cal Z}}

\newcommand{\todo}[1]{{\em \small {#1}}\marginpar{$\Longleftarrow$}}
\newcommand{\labell}[1]{\label{#1}\qquad_{#1}} 
\newcommand{\bbibitem}[1]{\bibitem{#1}\marginpar{#1}}
\newcommand{\llabel}[1]{\label{#1}\marginpar{#1}}
\newcommand{\PSbox}[3]{\mbox{\rule{0in}{#3}}}
\newcommand{\onefigurenocap}[1]{\begin{figure}[h]
         \begin{center}\leavevmode\epsfbox{#1.eps}\end{center}
         \end{figure}}
\newcommand{\onefigure}[2]{\begin{figure}[htbp]
         \caption{#2\label{#1}(#1)}
         \end{figure}}

\newcommand{\sphere}[0]{{\rm S}^3}
\newcommand{\su}[0]{{\rm SU(2)}}
\newcommand{\so}[0]{{\rm SO(4)}}
\newcommand{\bK}[0]{{\bf K}}
\newcommand{\bL}[0]{{\bf L}}
\newcommand{\bR}[0]{{\bf R}}
\newcommand{\tK}[0]{\tilde{K}}
\newcommand{\tL}[0]{\bar{L}}
\newcommand{\tR}[0]{\tilde{R}}

\newcommand{\btzm}[0]{BTZ$_{\rm M}$}
\newcommand{\ads}[1]{{\rm AdS}_{#1}}
\newcommand{\ds}[1]{{\rm dS}_{#1}}
\newcommand{\eds}[1]{{\rm EdS}_{#1}}
\newcommand{\sph}[1]{{\rm S}^{#1}}
\newcommand{\gn}[0]{G_N}
\newcommand{\SL}[0]{{\rm SL}(2,R)}
\newcommand{\cosm}[0]{R}
\newcommand{\hdim}[0]{\bar{h}}
\newcommand{\bw}[0]{\bar{w}}
\newcommand{\bz}[0]{\bar{z}}
\newcommand{\be}{\begin{equation}}
\newcommand{\ee}{\end{equation}}
\newcommand{\bea}{\begin{eqnarray}}
\newcommand{\eea}{\end{eqnarray}}
\newcommand{\pat}{\partial}
\newcommand{\lp}{\lambda_+}
\newcommand{\bx}{ {\bf x}}
\newcommand{\bk}{{\bf k}}
\newcommand{\bb}{{\bf b}}
\newcommand{\BB}{{\bf B}}
\newcommand{\tp}{\tilde{\phi}}
\hyphenation{Min-kow-ski}

\newcommand{\subdet}[1]{{\rm det}_{#1}}
\newcommand{\mydet}[0]{{\rm det}}

\def\apr{\alpha'}
\def\str{{str}}
\def\lstr{\ell_\str}
\def\gstr{g_\str}
\def\Mstr{M_\str}
\def\lpl{\ell_{pl}}
\def\Mpl{M_{pl}}
\def\varep{\varepsilon}
\def\del{\nabla}
\def\grad{\nabla}
\def\tr{\hbox{tr}}
\def\perp{\bot}
\def\half{\frac{1}{2}}
\def\p{\partial}
\def\perp{\bot}
\def\eps{\epsilon}
\newcommand{\Tr}{\mathop{\rm Tr}}

\renewcommand{\thepage}{\arabic{page}}
\setcounter{page}{1}


\def\NN{{\cal N}}
\def\nfour{{\cal N}=4}
\def\ntwo{{\cal N}=2}
\def\none{{\cal N}=1}
\def\nonestar{{\cal N}=1$^*$}
\def\tr{{\rm tr\ }}
\def\RR{{\cal R}}
\def\PP{{\cal P}}
\def\ZZ{{\cal Z}}

\newcommand{\bel}[1]{\be\label{#1}}
\def\al{\alpha}
\def\bt{\beta}
\def\mn{{\mu\nu}}
\newcommand{\rep}[1]{{\bf #1}}
\newcommand{\vev}[1]{\langle#1\rangle}
\def\bra{\langle}
\def\ket{\rangle}
\def\eref{(?FIX?)}

\rightline{UPR-T-943}
\rightline{WIS/15/01-JULY-DPP}
\rightline{hep-th/0107119}
\vskip 1cm
\centerline{\Large \bf Giant Gravitons in Conformal Field Theory}
\vskip 1cm

\renewcommand{\thefootnote}{\fnsymbol{footnote}}
\centerline{{\bf Vijay
Balasubramanian${}^{1,2}$\footnote{vijay@endive.hep.upenn.edu},
Micha Berkooz${}^{2}$\footnote{berkooz@clever.weizmann.ac.il},
Asad Naqvi${}^{1}$\footnote{naqvi@rutabaga.hep.upenn.edu},
and
Matthew J. Strassler${}^{1}$\footnote{strasslr@sage.hep.upenn.edu}
}}
\vskip .5cm
\centerline{${}^1$\it David Rittenhouse Laboratories, University of
Pennsylvania, Philadelphia, PA 19104, U.S.A.}
\centerline{${}^2$\it Faculty of Physics, The Weizmann Institute of
Science, Rehovot 76100, Israel.}
\vskip .5cm

\setcounter{footnote}{0}
\renewcommand{\thefootnote}{\arabic{footnote}}

\begin{abstract}
Giant gravitons in $\ads{5} \times \sph{5}$, and its orbifolds, have a
dual field theory representation as states created by chiral primary
operators.  We argue that these operators are not single-trace
operators in the conformal field theory, but rather are determinants
and subdeterminants of scalar fields; the stringy exclusion principle
applies to these operators.  Evidence for this identification comes
from three sources: (a) topological considerations in orbifolds, (b)
computation of protected correlators using free field theory and (c) a
Matrix model argument.  The last argument applies to $\ads{7} \times
\sph{4}$ and the dual $(2,0)$ theory, where we use algebraic
aspects of the fuzzy 4-sphere to compute the expectation value of a
giant graviton operator along the Coulomb branch of the theory.
\end{abstract}

\newpage

\section{Introduction}
\label{intro}
There is a remarkable mechanism in string theory for regulating
potential divergences and singularities: increasing the energy of a
state often causes it to expand, thereby softening its interactions. 
An example of this effect is the ``giant graviton''~\cite{suss} of the
$\ads{5} \times \sph{5}$ solution of IIB supergravity.  Here the Myers
effect~\cite{myerseffect} causes a graviton having very large angular
momentum on an $\sph{5}$ with nonzero Ramond-Ramond five-form flux to
expand into a spherical D3-brane, whose size is related to its angular
momentum.  This phenomenon suggests that one of the basic symmetries
of nature --- reparametrization invariance --- has a little-understood
generalization in the presence of Ramond-Ramond fluxes.  Apparently,
at very high energies in some backgrounds, the graviton, the ``gauge
boson'' of reparametrization invariance, becomes a macroscopic
extended object.  In view of this, it is clearly important to study
the properties of giant gravitons and their representation in dual
field theories.

$\ads{5}\times\sph{5}$ is dual to to ${\cal N}=4$ SYM \cite{mald}.
The standard map between these dual descriptions identifies single
giant gravitons with single-trace chiral operators of large R-charge
in the field theory.  The evidence for this identification, besides
the matching of charges, is two-fold.  First, at low R-charge, the
number of traces in an operator counts particle number in the dual
gravity Fock space.  If we identify a giant D3-brane as a single
graviton, then it should be created by a single-trace operator.
Second, there is a bound on the R-charge of single-trace operators in
the Yang-Mills theory.  This neatly matches the bound on the angular
momentum of a giant graviton, which arises because a spherical
D3-brane on $\sph{5}$ has a size that is bounded by the $\sph{5}$
radius.\footnote{This is the $\ads{5} \times \sph{5}$ analog of the
stringy exclusion principle in $\ads{3} \times \sph{3}$~\cite{exclusion}.}

In this paper, we show that the identification of giant gravitons as
single-trace operators in the dual field theory cannot be 
correct.\footnote{We will be interested in giant gravitons expanding 
on $\sph{5}$ rather than on AdS~\cite{myers,itsAki}.}
Rather, we will present evidence that large giant gravitons are dual
to states created by a family of subdeterminants.  (A subdeterminant of
an $N \times N$ matrix $X^{a}_{b}$ is
\begin{equation}
\det{}_{k} X = {1 \over k!}
\epsilon_{i_{1}\cdots i_{k}a_{1}\cdots a_{N-k}} \epsilon^{j_{1}\cdots
j_{k}a_{1}\cdots a_{N-k}} X^{i_{1}}_{j_{1}} \cdots
X^{i_{k}}_{j_{k}} \, .
\nonumber
\end{equation}
So $\subdet{N}$ is the same as the
determinant.)  These subdeterminants have a bounded R-charge, with
the full determinant saturating the bound.  Once again, the bound on
the R-charge is the field theory explanation of the angular momentum
bound for giants.

In Sec.~\ref{nottrace} we explain how the failure of the planar
approximation for correlation functions of large dimension operators
invalidates the single particle/single trace correspondence.  This
makes problematic the identification of giant gravitons as
single-trace operators.  In Sec.~\ref{orbifold} we identify
maximal-sized giant gravitons with determinants, using evidence from
various orbifolds of ${\cal N} = 4$ Yang-Mills theory and
corresponding dual gravities. In Sec.~\ref{subdet} we argue that
less-than-maximal giant gravitons are created by subdeterminants.
Sec.~\ref{matrix} presents a Matrix theory argument which shows that
giant gravitons in $\ads{7} \times \sph{4}$ are created by the analog
of subdeterminants in the $(2,0)$ theory, by relating dynamical
aspects of giant gravitons to algebraic aspects of the fuzzy 4-sphere.
We conclude, in Sec.~\ref{discussion}, with comments on future
directions.

Parts of Sec.~\ref{matrix} were developed in collaboration with Moshe
Rozali.

\section{Giant gravitons are not traces}
\label{nottrace}

The best semiclassical description of a graviton with large momentum
on the $\sph{5}$ of $\ads{5} \times \sph{5}$ is in terms of a large
D3-brane wrapping a 3-sphere and moving with some velocity\cite{suss}.
This is the giant graviton.  The transition from a graviton mode in a
spherical harmonic to a macroscopic brane is explained by the Myers
effect~\cite{myerseffect} in the presence of flux on the $\sph{5}$.
In $\ads{5} \times \sph{5}$, the radius of the spherical D3-brane is
$\rho^2=lR^2/N$, where $l$ is the angular momentum on the $\sph{5}$ of the
state, $R$ is the radius of the sphere, and $N$ the total 5-form flux
through the 5-sphere.

Our first point is that giant gravitons in the bulk supergravity
description are not created by single-trace operators in the conformal
field theory, as usually stated.  There are strong arguments that a
graviton with small angular momentum is created by a single-trace
operator, and that a state with $p$ gravitons is created by a product
of $p$ single-trace operators.  These arguments rely on the special
factorization properties of correlation functions in large-$N$ gauge
theories.  But factorization breaks down for operators with dimension
of order $N$ --- and thus for chiral operators with R-charge of order
$N$ --- so there are no such arguments for states containing giant
gravitons.

For example, consider two states with equal total angular momentum and
energy, but with that momentum and energy partitioned differently
among one or more giant gravitons.  These states are classically
different, and should represent orthogonal states in the Hilbert space
of the theory.  However, candidate multi-trace operators with the
correct charges do not generate orthogonal states (at least, the
familiar arguments of approximate orthogonality fail).

\subsection{Some basic tools}
\label{tools}
In this paper we will identify field theory operators that correspond
to giant gravitons on the gravity side of the AdS/CFT correspondence.
We will write these composite operators in terms of the fundamental
fields in the Lagrangian, a description that is useful at weak
coupling.  Strictly speaking the AdS/CFT correspondence applies at
strong field theory coupling and so the operators we identify should
be extrapolated to this regime.  Since they are chiral, there are no
significant problems with this extrapolation.  We will discuss two
types of systems: ${\cal N} =4$ theories, and nonperturbative fixed
points of ${\cal N} = 1$ theories (which can be constructed as
infrared fixed points of weakly-coupled theories.)

\paragraph{${\cal N}=4$ theories: }

The chiral primary operators of 4d, $\CN=4$, $SU(N)$ Yang-Mills theory
are in the $(0,l,0)$ representation of the $SU(4) \sim SO(6)$
R-symmetry group.  These operators, which may have one or more traces
over gauge indices, have protected dimensions $l\geq2$.  Single-trace
chiral primaries are of the form $\Tr \,(X^{i_1}X^{i_2} \cdots
X^{i_l})$, where each scalar $X^i$ is a vector of SO(6), while the
product of scalars forms a symmetric traceless representation of this
group.  These operators generate the chiral ``ring'' of the theory. 
There is a bound $l\leq N$ for the single-trace chiral primaries,
because all higher $l$ operators can be written as sums of products of
lower dimension operators.  

We are also interested in the $SO(2N)$, ${\cal N}=4$ theory. The
spectrum of single-trace operators is similar, except that only the
$(0,2l,0)$ representations appear, and there is one additional
representation needed to generate the chiral ring.   The
corresponding operator is a Pfaffian of the scalar fields,
which are antisymmetric $2N\times 2N$ matrices.

In $\CN=1$ supersymmetric language, the $\CN=4$ theory consists of a vector
multiplet and three chiral multiplets $\Phi_\mu=X_\mu+iX_{\mu+3}$,
$\mu=1,2,3$, all in the adjoint representation of the gauge group.
The $SO(6)$ R-symmetry is partially hidden by the $\CN=1$ notation.
Only a $U(1)$ R-symmetry and the $SU(3)$ that rotates the $\Phi_\mu$
are visible.  The chiral operators described above include operators
of the form $\Tr(\Phi^{\mu_1}\Phi^{\mu_2}\cdots \Phi^{\mu_n})$
(symmetrized over the $SU(3)$ indices).  However, there are chiral
operators of the $\CN=4$ theory which appear to be non-chiral in the
$\CN=1$ notation.  This is because a short representation of the
$\CN=4$ algebra includes both short and long representations of a
$\CN=1$ subalgebra.  In the rest of this paper, we will use the
$\CN=1$ language.  Also, for simplicity, we will examine
R-charges in a single $U(1)$ and will suppress the index $\mu$.

In these theories $g_{ym}^2$ is a marginal coupling constant (and it
couples to some operator $O_{marg}$). We will only be studying chiral
operators
\footnote{We define these at weak coupling using the fundamental
fields in a weakly coupled Lagrangian. To define the operators 
in  the strong coupling region, a natural requirement would be that
\begin{equation}
\partial_g\langle{\hat O}\cdots {\hat O}\rangle_g = \langle{\hat
O}_1^{(g)}\cdots {\hat O}_k^{(g)} \int d^4x \, O_{marg}\rangle_g
\label{partras}
\end{equation} where $\langle\rangle_g$ denotes evaluating the
expectation value at some value $g$ of the coupling.}  and extremal
correlation functions, for which there are powerful
non-renormalization theorems. If $O_p$
is any scalar chiral primary operator in the $(0,l,0)$ representation
(whether a single trace or a multi-trace operator), then extremal
correlators are correlators of the form
\begin{equation}
<O_{p_1}\cdots O_{p_n}>,\ \ \ p_1=p_2+\cdots p_n.
\end{equation}
These are independent of the coupling.\footnote{One can also define
near-extremal correlators, which satisfy $p_1>p_2+\cdots p_n-k$ for
small $k$. Some non-renormalization theorems apply for such
correlators as well.}. The result is that we can compute these
correlators reliably at weak coupling.

For the most part we will use an even more restricted class of
correlators which are just 3 point functions of chiral primaries. In
this case all such 3 point functions, irrespectively of whether they
are extremal or not, are not renormalized \footnote{We are grateful to
Massimo Bianchi, Dan Freedman and Kostas Skenderis for emphasizing
this to us.}.

These non-renormalization theorems were first discussed in
\cite{nonrena}, extended in \cite{nonrenb}, extensively checked by
supergravity and field theory computations in \cite{nonrenc}, and was
finally proven in \cite{nonrend}. Whereas most of the references
\cite{nonrena,nonrenb,nonrenc} focussed on single-trace operators,
\cite{nonrend} relies, as a superspace proof has to, only on algebraic
properties of the operators.  Hence it applies to all primaries
operators in chiral $(0,l,0)$ representations, whether they are
defined as single-trace at weak coupling or not.

We will also be referring to finite ${\cal N}=2$ and ${\cal N}=1$
theories. In the ${\cal N}=2$ case, the non-renormalization theorem
applies just as well \cite{nonrend}. In the ${\cal N}=1$ case, there
is no exact non-renormalization theorem, although one expects some
remnant of it at large $N$ for theories obtained by orbifolding ${\cal
N}=2$ or $4$ theories.

\paragraph{Fixed points of ${\cal N}=1$ theories: }

We will also encounter ${\cal N}=1$ fixed point theories that can be
defined by starting with a weakly-coupled Lagrangian description in
the UV and flowing to a strongly-coupled fixed point in the IR. The
operators are defined in the UV using fundamental fields in the
Lagrangian, and the IR objects are particles or branes on $AdS\times
X$.  In this case, there is no non-renormalization theorem for
correlation functions and operators mix.  Nevertheless, we will
consider operators corresponding to topologically stable objects on
$AdS\times X$.  Such operators are the lowest dimension objects
carrying certain discrete quantum numbers.  Hence their operator
mixing is extremely limited.  We will also consider products of
operators of this type.  Since these are products of chiral operators
no new renormalizations or operator mixings are introduced.

\subsection{Small angular momenta}

According to the standard AdS/CFT dictionary, states created by the
chiral primary operators of the $\CN =4$, $SU(N)$ theory (more
precisely, generators of the chiral ring) map into modes of fields of
IIB supergravity in the background $\ads{5} \times \sph{5}$.  The
SO(6) isometries of the sphere appear as R-symmetries of the AdS
theory as well as the dual CFT. Spherical harmonics on the sphere
lead, upon Kaluza-Klein reduction to AdS, to a tower of massive fields
transforming in the $(0,l,0)$ representations of SO(6).  These fields
have a Fock space of single and multi-particle states in AdS which
form an orthogonal basis.  In the standard AdS/CFT map, particle
number in AdS is mapped into the number of traces in the operator
creating the dual field theory state.  It is instructive to examine
the rationale behind the identification of these quantum numbers.

\begin{figure}
  \begin{center}
 \epsfysize=2in
   \mbox{\epsfbox{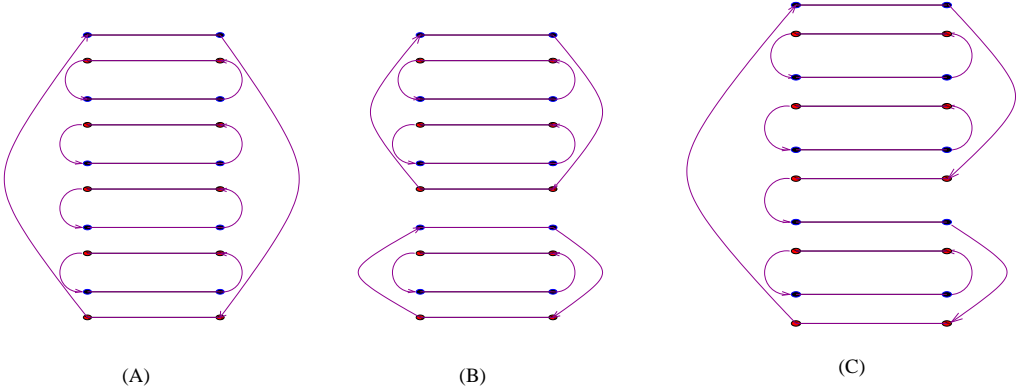}}
\end{center}
\caption{ Planar diagrams for :(A)
$\langle \CO_1 \CO_1 \rangle$ (there are $l$ such diagrams from
cyclic permutations inside the Trace), (B)
 $\langle \CO_2 \CO_2 \rangle$ (there are $l_1\, l_2$ such diagrams),
(C)  $\langle \CO_1 \CO_2 \rangle$ (there are $l_1\, l_2\, l$ such
diagrams).}
\label{fig:norm}
\end{figure}

In general, a state created by a composite operator like $\Tr  X^l$
need not be orthogonal to states created by other gauge invariant
operators with the same dimension.   For example,
\begin{equation}
\langle  \Tr( \Phi^l)^\dagger \, (\Tr (\Phi^{l -2}) \, \Tr \Phi^2 ) \rangle
\neq 0
\end{equation}
However, if the rank of the gauge group $N$ is large compared to $l$,
there is approximate orthogonality because of well-known properties of
this limit.  For example,  let
\begin{eqnarray}
\CO_1 & = &  \Tr (\Phi^l) \, , \label{op1} \\
\CO_2 & = &  \Tr \Phi^{l_1} \Tr
\Phi^{l_2}\,\,\,\,\,\,\,\,\,\,\,\,\,\,\,\,\,
l_1+l_2=l
\label{op2}
\end{eqnarray}
Then,
\[
\langle \, \CO_1^{\dagger} \, \CO_1  \, \rangle \sim lN^l,
\]
\[
\langle \, \CO_2^{\dagger} \, \CO_2 \, \rangle \sim l_1 l_2 N^l
\]
where we have left out the space-time dependence.  To compute this
2-point function in the large-$N$ limit, we have recognized that
planar diagrams dominate and have done the $N$-counting by looking
at the free diagrams in Fig.~\ref{fig:norm}A and B.
With these results we can compute the normalized 2-point function of
two different operators in the planar approximation.  The $N$-counting
can again be understood from the free diagram in Fig.~\ref{fig:norm}C:
\begin{equation}
\frac{\langle \, \CO_1^{\dagger} \, \CO_2 \, \rangle}{\sqrt{\langle \,
\CO_1^{\dagger} \, \CO_1 \, \rangle}\sqrt{{\langle \, \CO_2^{\dagger}
\, \CO_2 \, \rangle}}} \sim \frac{\sqrt{l_1\, l_2\, l}} {N}
\label{ortho}
\end{equation}
By the state-operator correspondence in CFTs, this implies that
the states created by $\CO_1$ and $\CO_2$ are orthogonal in the $N
\rightarrow \infty$ limit.  In general, operators with the same
dimension but with different numbers of traces create states that
are orthogonal at large $N$, as in (\ref{ortho}).  This suggests
that the number of traces is a conserved quantum number at large
$N$. Furthermore, disconnected graphs dominate in two-point
functions of multi-trace operators, so it appears that a product
of single-trace operators gives a state which is a direct product
of noninteracting single particle states.  Altogether, this leads
to the conclusion that the number of particles in AdS should be
mapped into the number of traces in the CFT.

\subsection{Failure of planarity and orthogonality of traces}
\label{failplane}

It is well known that the dominance of planar graphs at large $N$ is
only true for operators whose dimensions are small.  For correlation
functions of operators whose dimensions scale sufficiently fast with
$N$, it fails badly \cite{wtnbar}.  It is readily checked that when
$l$ is sufficiently large, combinatorial factors in non-planar
diagrams can overwhelm the usual suppression by powers of $1/N^{2}$. 
When $l$ is large, even the planar calculation (\ref{ortho}) does not
show orthogonality of states created by operators with different
numbers of traces.  For example, (\ref{ortho}) shows that the planar
contribution to the two-point function scales as:
\begin{eqnarray}
    l = O(N), \,\, l_{1} = O(N),\,\, l_{2} = O(1)
    ~~~~&\longrightarrow&~~~~
    \frac{\langle \, \CO_1^{\dagger} \, \CO_2 \, \rangle}{\sqrt{\langle \,
\CO_1^{\dagger} \, \CO_1 \, \rangle}\sqrt{{\langle \, \CO_2^{\dagger}
\, \CO_2 \, \rangle}}} \sim O(1)  \label{2pt1} \\
 l \sim N, \,\, l_{1} = l_{2} = l/2
    ~~~~&\longrightarrow&~~~~
    \frac{\langle \, \CO_1^{\dagger} \, \CO_2 \, \rangle}{\sqrt{\langle \,
\CO_1^{\dagger} \, \CO_1 \, \rangle}\sqrt{{\langle \, \CO_2^{\dagger}
\, \CO_2 \, \rangle}}} \sim \sqrt{N}  \label{2ptgrow}
\end{eqnarray}
It is rather unlikely that the large non-planar contributions to this
two-point function will conspire to cause it to cancel at large $N$,
although they should correct the apparently non-unitary growth of
(\ref{2ptgrow}), in conjunction with non-perturbative effects.  The
salient point is that although it is hard to do an exact computation
of the two-point function even in free field theory, there is no
reason to expect that states created by high dimension operators with
different numbers of traces are orthogonal.  ``Trace number'' is
simply not a good quantum number.  A consequence of this for the
AdS/CFT duality is that we cannot map CFT trace number into particle
number in an AdS Fock space for states of high dimension.

\subsection{Scales in gravity, gauge theory and giants}

The ground state wavefunction for a scalar field on $\ads{5}$
arising from a mode with large angular momentum $l$ on $\sph{5}$
is~\cite{bkl}:
\begin{equation}
\psi \propto e^{-i \,l \, t} {1 \over (1 + r^2/R^2)^{l/2}}
\end{equation}
Here $r$ is a radial coordinate in AdS, $R = (g_s N \alpha'^2)^{1/4}$
is the curvature scale, and $l$ is the angular momentum on $\sph{5}$. 
The effective mass of the scalar field, induced by the large angular
momentum, is $(mR)^{2} \sim l^{2}$.  The typical size of this
wavefunction decreases as $l$ increases, so that when $l > (g_s
N)^{1/2}$ it has a width smaller than the string scale and therefore
cannot be trusted.  Equivalently, semiclassical supergravity
calculations can only trusted for particles with masses below the
string scale ($l < (g_s N)^{1/2}$) and the Planck scale ($l<
N^{1/2}$).  Another criterion that can be applied is to require that
the wavelength of modes on the $\sph{5}$ is larger than the string
length or Planck length.  This gives bounds $l < (g_s N)^{1/4}$ and
$l< N^{1/4}$ respectively.  It is clear that for such small values of
$l$ the two-point function (\ref{ortho}) approaches zero, the planar
approximation is valid, and the standard mapping of one-particle
states to single-trace operators continues to hold.

In~\cite{suss}, it was argued that a large class of states with large
angular momenta should be described by spherical, BPS D3-branes which
are extended on $\sph{5}$ but localized at a point on $\ads{5}$.  Such
spherical brane configurations on an $\sph{5}$ of radius $R$ are
supported by the angular momentum $l$ and the $N$ units of background
flux.  They have a radius $\rho^{2} = l R^{2}/N$. Since $\rho$ cannot exceed
$R$, there is a bound on the angular momentum $l \leq N$.  The
semiclassical description of giants shows that there is a clear
distinction between single- and multi-giant states, and that, despite
being non-perturbative objects, they interact weakly with perturbative
gravitons and with each other.

A giant graviton can be treated as a non-perturbative state in weakly
coupled string theory when its size $\rho$ exceeds the Planck scale
($l_{p}^{8} = l_{s}^{8} g_{s}^{2}$).  After applying the AdS/CFT
dictionary, this requires that $l > N^{1/2}$.  The dual CFT two-point
function (\ref{ortho}) shows that the planar approximation for
correlation functions, and with it the argument for orthogonality of
multi-trace operators, breaks down at least when $l > N^{2/3}$.
Therefore, since single- and multi-giant states are clearly orthogonal
at weak coupling, large giants cannot be described by single- and
multi-trace chiral primaries of large R-charge.  Indeed, traces can be
a good description at most when $l < N^{2/3}$.  What is more, since
trace number is not a good quantum number at large $l$, the bound on
the R-charge of single trace chiral primaries cannot be mapped
meaningfully to the angular momentum bound for giant gravitons.

\paragraph{The Puzzle: }
Nevertheless, giant gravitons exist as semiclassical entities.  They
are non-perturbative, BPS brane states, and we expect configurations
of giants with the same total angular momentum but different numbers
of branes to be orthogonal.  Brane number, unlike particle number,
makes sense from the point of view of gravity at these high energies.
Although single- and multi-trace chiral primaries make states that
carry the same charges as the giant gravitons, we cannot map ``trace
number'' to brane number since the former is not a good quantum
number.  As a corollary, we cannot relate the bound on the angular
momentum of single giants to the bound on the charge of single-trace
operators.  What operator, then, makes a giant graviton in the CFT?
And whence the bound on the angular momentum?

\section{Biggest giants should be determinants}
\label{orbifold}

The resolution of these questions begins with the simple observation
that we have seen D3-branes in the AdS/CFT correspondence before.
Specifically, topologically {\it non-trivial} D3-branes play an
important role in theories in which the five-sphere is orbifolded, or
replaced with another space with non-trivial
three-cycles, such as $T^{1,1}$.

A BPS D3-brane wrapped on such a 3-cycle corresponds to an operator in
the CFT of dimension $\sim N$.  Such a state carries a topological
charge in addition to its angular momentum, and is the lightest object
with this charge.  Unlike non-topological BPS gravitons, which are
neutrally stable with respect to states with multiple BPS gravitons,
these wrapped D3-branes cannot split into multi-brane or
brane-graviton states.  The corresponding chiral operator in the CFT
is therefore easy to identify, since it must be the lowest-dimension
chiral operator with a certain conserved charge corresponding to the
wrapping charge of the brane.  In many cases this charge is a baryon
number, and the operator is baryon-like.

  In each case, there are also chiral operators with the cancelling
topological charge.  A product of chiral operators, each of which
carries topological charge but whose combination is neutral, is
naturally associated to a combination of wrapped D3-branes each
of which is topologically nontrivial but which when joined together
are topologically trivial.  Such D3-branes are in the same class as
the giant gravitons of~\cite{suss}, which are topologically-trivial BPS
brane states that are dynamically stabilized by their angular
momentum.  As we will see, these products of topologically-charged
chiral operators are always determinants of neutral combinations of
the matrix-valued fields in the CFT.  These determiniants carry the
maximum angular momentum on the space, and are therefore candidates
for the maximal giant gravitons, those at the edge of the angular
momentum bound.

We begin with the most carefully studied case, the near-horizon
region near $N$ D3-branes on a ${Z_{3}}$ orbifold.

\newcommand{\z}[0]{Z_{3}}
\subsection{The $\z$ orbifold}

\paragraph{Gravity: }
 The sphere in the $\ads{5} \times \sph{5}$
solution of IIB gravity can be parametrized in complex coordinates
$(z_{1},z_{2},z_{3})$ as the solution to $|z_{1}|^{2} + |z_{2}|^{2} +
|z_{3}|^{2} = R^{2}$.  Consider the $\z$ orbifold:\footnote{We will
follow the discussion of the $\z$ orbifold in~\cite{rgw}.}
\begin{equation}
    z_{i} \sim z_{i} \, \omega  ~~~~;~~~~ \omega = e^{2\pi i /3}\,.
\end{equation}
This orbifold has no fixed points, and has non-trivial 3-cycles on
which D3-branes can be wrapped. Specifically,
\begin{equation}
    H_{3}(\sph{5}/\z,Z) = \z \, .
\end{equation}
This torsion class is generated by an $\sph{3}/\z$ which is defined,
up to $U(3)$ transformations of the $z_{i}$, by $z_{3} = 0$.  Branes
wrapped on such cycles carry a $\z$ valued D3-brane charge.

The first integral homology group of the orbifold is
\begin{equation}
H_{1}(\sph{5}/\z,Z) =
\z \, .
\end{equation}
This torsion class is generated by an $\sph{1}/\z$ given, up to $U(3)$
transformations, by $z_{2} = z_{3} = 0$.  Fundamental strings and
D-strings can be wrapped on this cycle giving twisted sectors with a
$\z$ valued charge.  The same cycle leads to a non-trivial fundamental
group $\pi_{1}(\sph{3}/\z) = \z$ on the wrapped D3-brane worldvolume.

Discrete Wilson lines can be chosen on the worldvolumes of branes
wrapped on cycles with nontrivial $\pi_{1}$, leading to a set of
D-brane bound states \cite{gopvaf}.  In the present case, where
$\pi_{1} = \z$ is discrete, Abelian and finite, there are Wilson
lines, with three possible monodromies $1$, $\omega$ and $\omega^{2}$
around the nontrivial 1-cycle.  By turning on such Wilson lines, we get
three discrete vacua for a 3-brane wrapped once around $\sph{3}/\z$.  We
will call these states, which are cyclically permuted by the quantum
$\z$ symmetry of the orbifold, $| U \rangle$, $|V \rangle$ and $|W
\rangle$.  Following~\cite{gopvaf}, we can take linear combinations of
these vacua to get eigenstates of the quantum $\z$ which carry
definite fundamental string charge:
\begin{equation}
|U\rangle + |V\rangle + |W\rangle ~~~;~~~
|U\rangle + \omega \, |V\rangle + \omega^{2} \, |W\rangle ~~~;~~~
|U\rangle + \omega^{2} \, |V\rangle + \omega \, |W\rangle \, .
\end{equation}
These vacua are bound states of the wrapped D3-brane
with a twisted sector string.  Conversely, the states $| U \rangle$,
$|V \rangle$ and $|W \rangle$ carry indefinite F-string charge, but
carry definite D1-brane charge in accord with the non-commutation of `t
Hooft and Wilson lines, as discussed in~\cite{rgw}.

We are interested in the largest dynamically stable giant graviton. 
For concreteness, consider only operators which are charged under a
$U(1)$ part of the global $U(3)$ symmetry which rotates just the $z_1$
coordinate.  The corresponding giant gravitons are D3-branes which
wrap an $\sph3$ (${|z_2|}^2+{|z_3|}^2=\rho^2$), and rotate along the
circle ${|z_1|}^2=R^2-\rho^2$.  The radial size of the spherical brane
is determined by $\rho^2=lR^2/N$~\cite{suss}, where $R$ is the radius
of the $\sph5/Z_3$.  On the covering space ($\sph5$), we can consider
three branes at image points under $Z_3$.  To reach the maximal size
we take $l\rightarrow N$, getting three D3-branes at $z_1=0$.  On the
covering space these three branes wrap the 3-sphere $|z_{1}|^{2} +
|z_{2}|^{2} = R^{2}$, which is modded out by the action of the $\z$ to
become an $\sph{3}/\z$ in the orbifold.  To study maximal giant
gravitons we should therefore examine topologically trivial states
that are triply-wrapped on the $\sph{3}/\z$ inside $\sph{5}/\z$.

A D3-brane that is triply-wrapped on $\sph{3}/\z$ is topologically
trivial and should therefore carry no charge under the quantum $\z$
that permutes $|U\rangle$, $|V\rangle$ and $|W\rangle$.  We can
construct such a state by placing all three kinds of topologically
protected branes that were discussed above on the $\sph{3}/\z$ at the
same time.  The resulting state, which we will call $|UVW\rangle$, has
a $U(3)$ Wilson line with monodromy ${\rm diag}(1,\omega,\omega^{3})$.
This can be written, after a change of basis, as the permutation
matrix
\begin{equation}
    \pmatrix{0 & 1 & 0 \cr 0 & 0 & 1 \cr 1 & 0 & 0 }
    \, ,
\end{equation}
as appropriate to a triply-wrapped brane.\footnote{The states
$|UUU\rangle$, $|VVV\rangle$ and $|WWW\rangle$ are topologically
charged.  The linear combination $|UUU\rangle + |VVV\rangle +
|WWW\rangle$ is not charged, but describes a $U(3)$ Wilson line which
is proportional to the identity and has an indefinite overall phase. 
This therefore represents three individual wrapped branes with an
indefinite overall phase, rather than a topologically trivial
triply-wrapped brane.}

Although such a brane is topologically trivial, it is stabilized by
dynamical effects.  This can be seen by examining the arguments
in~\cite{suss} that originally established the existence of giant
gravitons.  Those authors used the Born-Infeld action to demonstrate
that a spherical brane with an angular momentum $l$ on $\sph{5}$ is
stabilized by the flux.  The maximal giant graviton (of radius $R$) in
their case is actually stationary and acquires all its angular
momentum from interaction with the flux via the Chern-Simons term
in its action.  The Born-Infeld calculation of~\cite{suss} can be
repeated in our case by observing that it is locally unchanged on the
worldvolume of the orbifolded brane.  Therefore, the BPS condition
relating energy and charge follows automatically.\footnote{It is
easiest to compute the Hamiltonian for the maximal giant
in~\cite{suss} and apply that directly to our setting, rather than
relying on their Lagrangian which develops coordinate singularities in
the limit of the maximal giant graviton.}

The dynamically stabilized state that we have constructed is a maximal
size giant graviton of the orbifold theory, since the three $\sph{3}/\z$
surfaces from which it is constructed have the maximum radius ($R$)
that a three cycle can have on the $\sph{5}/\z$.   We will show how
this giant is described in the dual field theory.

\paragraph{Field Theory: } $\ads{5} \times \sph{5}/\z$ is the near
horizon limit of $3N$ D3-branes placed at the fixed point of ${\cal
C}^{3}/\z$~\cite{rgw,evashamit}.  The resulting $\CN = 1$ field theory
on the D-branes has a gauge group $SU(N) \times SU(N) \times SU(N)$.
There are chiral matter superfields $U_{\mu}$, $V_{\mu}$ and
$W_{\mu}$, $\mu=1,2,3$ each transforming as a bifundamental of a pair
of these $SU(N)$ factors and as a singlet of the third.  The theory
has a $U(1)$ $\CN=1$ R-symmetry, and an ordinary $SU(3)$ global
symmetry which rotates the index $\mu$.  This $U(3)$ is the subgroup
of the $SO(6)$ $\CN = 4$ R-symmetry which survives the orbifold
projection; it rotates the complex coordinates $z_{i}$ of ${\cal
C}^{3}/\z$.

Detailed identification of the states and symmetries of this theory
with string theory on $\ads{5} \times \sph{5}/\z$ was carried out
in~\cite{rgw}.  These authors argue that the D3-brane states that
are singly-wrapped on $\sph{3}/\z$ --- $|U\rangle$, $|V\rangle$ and
$|W\rangle$ --- are created from the CFT vacuum by the gauge invariant
``dibaryon'' operators
\begin{equation}
\det U, \, \,  \det V,  \, \, \det W
\end{equation}
where
\begin{equation}
    \det U = \epsilon_{i_{1}\cdots i_{N}} \,
    \epsilon^{j_{1}\cdots j_{N}} \,
     U^{i_{1}}_{j_{1}} \cdots  U^{i_{N}}_{j_{N}} \, .
     \label{detdef1}
\end{equation}
Note that $i_s$ and $j_s$ are indices in different $SU(N)$ factors. 
We have left out the $U(3)$ index $\mu$.  Appropriate symmetrization
over the latter index would be needed to construct different $U(3)$
representations, but we will not need to consider the most general
giant; we can instead take all $\mu=1$.

 The maximal giant graviton of the orbifold
theory is the triply-wrapped brane state $|UVW\rangle$.  This is
created by the operator
\begin{equation}
    \det U \, \det V \, \det W \equiv \det(UVW)
= \epsilon_{i_{1}\cdots i_{N}} \,
    \epsilon^{{\bar i}_{1}\cdots {\bar i}_{N}} \,
     (UVW)^{i_{1}}_{{\bar i}_{1}} \cdots  (UVW)^{i_{N}}_{{\bar i}_{N}} \, .
     \label{detdef2} \, .
\end{equation}
Note that $i_s$ and ${\bar i}_s$ are in the {\it same} $SU(N)$ factor.
Thus, the BPS D3-brane wrapped on the largest possible non-topological
3-cycle is created in the dual CFT by a determinant of a
topologically-neutral combination of matrix-valued fields.

\subsection{Other examples}
\paragraph{SO(2N): }
The $SO(2N)$, $\CN = 4$ Yang-Mills theory is dual to $\ads{5} \times
\RR \PP^5$.  The theory has six scalar fields $X_i$ in the adjoint
representation; these are antisymmetric tensors in color space. 
Consider the complex combination $\Phi = X_1 + i X_2$.  There is a
gauge-invariant, symmetric combination of $N$ of these tensors with a
single $2N$-index epsilon tensor.  This is the Pfaffian of $\Phi$,
${\rm Pf} (\Phi)$, which was identified by \cite{wtnbar} as creating
the state with a single BPS D3-brane wrapped on the nontrivial
three-cycle of $\RR\PP^5$, and carrying a conserved $Z_2$ quantum
number.  The maximal giant graviton is the topologically trivial brane
state which wraps twice the non-trivial cycle.  Again, this can be
understood by taking a near-maximal giant graviton with the quantum
numbers of $\Phi^l$.  This is a D3-brane wrapping an $\sph3$ and
rotating at ${|\Phi|}^2=R^2(1-l/N)$.  As $l\rightarrow N$,
$\rho\rightarrow 0$ and we obtain a brane doubly wrapping the
$\RR\PP^3$ at $\Phi=0$.  This doubly wrapped D3-brane is topologically
unstable but is dynamically stable following~\cite{suss}.  Since
\begin{equation}
{\rm Pf} (\Phi) \, {\rm Pf} (\Phi)= \det \Phi \, ,
\end{equation}
we see that the maximal giant graviton of the $SO(2N)$ theory
should be identified with $\det\Phi$.

\paragraph{$Z_{2}$ orbifold: }
The finite $SU(N)\times SU(N)$, $\CN=2$ gauge theory with two
hypermultiplets $A_i, B_i$ ($i=1,2$) in the bifundamental
representation is dual to $\ads{5} \times \sph5/Z_2$~\cite{evashamit}. 
This $\CN =2$ theory has oriented three-cycles whose winding number
takes values in $Z$, and which have two possible orientations.  A
D3-brane wrapped on a cycle with one orientation is a dibaryon created
by $\det A$, suppressing global symmetry quantum numbers for the
moment.  The topological charge is dibaryon-number, and it is
integer-valued, as there is no way to reduce the operator $(\det A)^n$
to sums of other operators.  A wrapped D3-brane in the opposite
orientation is the diantibaryon $\det B$, carrying opposite
topological charge.  A di-baryon-antibaryon state is wrapped doubly
and non-topologically around the equator of $\sph5/Z_2$, but is
dynamically stabilized following the discussion above.  This is a
maximal size giant graviton of the orbifold theory.  Clearly, since
\begin{equation}
    \det A \, \det B = \det (AB) \, .
\end{equation}
the maximal giant graviton is again identified  with a determinant.

\paragraph{The conifold: }
There is a conformal $SU(N)\times SU(N)$, $\CN = 1$ gauge theory with
chiral multiplets $A_i, B_j$ ($i,j=1,2$) in the bifundamental and
a antibifundamental representation and with superpotential $W = \tr
(A_iB_jA_kB_\ell ) \epsilon^{ik}\epsilon^{j\ell}$.  This is dual to
$\ads{5} \times T^{1,1}$, the base of the conifold treated as a
cone~\cite{klebwit,klebgub}.  Like the $Z_{2}$ orbifold, $T^{1,1}$ has
3-cycles with integer winding numbers and two orientations.  D3-branes
wrapped in the two orientations give rise to dibaryons and
antidibaryons created by $\det A$ and $\det B$.  (Again, we have
suppressed global quantum numbers.)  In parallel, with the $Z_{2}$
orbifold, a di-baryon-antibaryon state is a topologically trivial
D3-brane wrapped around an equator of $T^{1,1}$.  This, state, created
by
\begin{equation}
\det A \, \det B = \det (AB) \, ,
\end{equation}
is a maximum size giant graviton, i.e., a BPS D3-brane
wrapped on the largest allowed nontopological 3-cycle.  Once again, it
is created by a determinant in the dual theory.

\subsection{The maximal giant on $\ads{5} \times \sph{5}$}

The examples given above strongly suggest that the maximal giant
graviton on $\ads{5} \times \sph{5}$, namely the largest spherical BPS
brane on $\sph{5}$, should be created in the $\CN = 4$ $SU(N)$
Yang-Mills theory by a gauge-invariant operator that has to do with
determinants.  The natural guess is
\begin{equation}
    O_{N} = \det \Phi = {1 \over N!}
      \epsilon_{i_{1}\cdots i_{N}}
\epsilon^{j_{1}\cdots j_{N}} \Phi^{i_{1}}_{j_{i}}
\Phi^{i_{N}}_{j_{N}} \, .
\end{equation}
(We have chosen an addition $1/N!$ in the normalization relative to
(\ref{detdef1}) for later convenience.)  In Sec.~\ref{matrix} we will
give a more direct justification for this guess by studying the 6d
${(2,0)}_k$ theory, which reduces to $d=4, {\cal N}=4$ $SU(N)$ when
compactified on $T^2$.

The determinant enjoys an
expansion in terms of traces,
\begin{equation}
\det \Phi= \sum_{
\{ k_r \}|\sum(k_r) = N} c_{\{k_r\}}\prod_{r} \tr \Phi^{2k_r}  \,  ,
\label{exp1}
\end{equation}
which is worked out in Appendix~\ref{detexp}.  The reader might wonder
if this expansion is dominated by the single-trace term, making our
arguments consistent with the usual discussion of giant
gravitons~\cite{suss}.  Indeed, Eq.  (\ref{dettraceeq}) in
Appendix~\ref{detexp} shows that the terms with multiple traces in
(\ref{exp1}) are somewhat suppressed by powers of $1/N$.  However,
there are two counts against this line of thinking.  The first is that
there are exponentially more multi-trace than single-trace terms.  As
a result, even if the states created by single and multiple trace
operators in (\ref{exp1}) were orthogonal, phase-space effects would
imply that the maximal giant graviton is better seen as a
superposition of multi-particle states than as a one-particle state
created by a single-trace chiral primary.  Still worse, however, as
explained in Sec.~\ref{failplane}, the states created by the dimension
$\Delta=N$ single- and multi-trace operators appearing in the
expansion of the determinant (\ref{exp1}) are not orthogonal (or at
least need not be orthogonal) to each other.  Consequently, the
overlap between the state created by $\det\Phi$ and a state created by
$\prod_{r} \tr \Phi^{k_r}$ is therefore not proportional to the
coefficient $c_{\{k_r\}}$ appearing in (\ref{exp1}).  The overlap
between the maximal single-trace and the determinant is unknown, but
there is no reason it should be of order one.\footnote{To carefully
check these assertions one needs to consistently normalize the
multi-trace operators in (\ref{exp1}).}

\paragraph{Summary: } We have shown that the largest giant graviton in
a variety of orbifold theories is represented in the dual field theory
by the determinant of a suitable combination of scalar fields.  For
example, the maximal giant of the $SO(2N)$ theory is created by
$\det\Phi$ where $\Phi$ is a scalar field of the theory.  We argued
that this strongly suggests the identification of the largest giant
graviton in $\ads{5} \times \sph{5}$ with the CFT state created by
$\CN = 4$, $SU(N)$ operator $O_{N} = \det\Phi$.  To complete the
identfication, we need a family of CFT operators $O_{l}$ that describe
giant gravitons with angular momentum $l$, and approach the
determinant $O_{N}$ as $l \rightarrow N$.

\section{Subdeterminants and the giant graviton bound}
\label{subdet}

In the previous section we identified maximal giant gravitons with
determinants in the dual CFT. The difference between the
topologically-stable D3-branes in the orbifold, orientifold and
conifold models and the topologically-trivial maximal giant gravitons
is that the latter can be continuously shrunk to zero size and angular
momentum by emission of ordinary gravitons.  There must therefore be a
class of operators in the CFTs whose dimension and angular momentum
can be gradually shrunk from $N$ to zero, with the highest element of
the class being the determinant operator.  However, this class must
not exist for the stable dibaryons and/or Pfaffians discussed above.

In this section, we will give two heuristic arguments that the natural class
of operators in the $\CN = 4$ $SU(N)$ theory is the set of subdeterminants
\begin{equation}
O_l=\det {}_l  \Phi \equiv \frac{1}{l!}\, \epsilon_{i_1 i_2 \cdots i_l a_1
a_2 \cdots
a_{N-l}}\,
\epsilon^{j_1 j_2 \cdots j_l a_1 a_2 \cdots
a_{N-l}}\, \Phi^{i_1}_{j_1}\, \Phi^{i_2}_{j_2}\,\cdots\, \Phi^{i_l}_{j_l}
\label{subdets}
\end{equation}
In the  $\z$ orbifold model,  operators of this type also exist (with each
$\Phi$ replaced by $UVW$) but no such gauge-invariant
generalizations of the determinant exist for the matrix $U$ alone.
Similar statements apply to the other cases; we will discuss
the $\CN = 4$ $SO(2N)$ in some detail.  Thus, these operators
are the natural candidates for non-maximal giant gravitons.  Like
the giants, they have a bound of $N$ on their mass and angular momentum.

If these operators create giant gravitons, then they must satisfy
orthogonality conditions.  We will show in this section, by complete
(nonplanar) evaluation of some of their correlation functions, that
they indeed create orthogonal states.  Moreover, we will see that the
probability of a giant graviton to emit a small graviton is small but
not highly suppressed, while the probability for a giant to split into
two giants is exponentially suppressed.

\subsection{A proposal for subdeterminants}

\paragraph{{\bf SU(N)}:} In the $\CN =4$, $SU(N)$ Yang-Mills theory we
have proposed that the operator $\det\Phi$ makes the maximal giant
graviton.  We expect that giants of size close to maximal will be
created by similar operators.  On the other hand we know that for
small angular momenta the operators $\Tr \Phi^{l}$ are a good CFT
representation of supergravity modes.  A simple heuristic method of deriving a
candidate set of operators for less-than-maximal giant gravitons is to
consider a process where a small perturbative graviton made by $\Tr
\Phi^{\dagger 2}$ is absorbed by a maximal giant made by $\det\Phi$. 
The would expect that the most likely result of such a collision would
be a smaller giant graviton.  We can study this in the field theory by
identifying the chiral $N-2$ R-charge operator in the OPE of
$\det\Phi$ and $\Tr \Phi^{\dagger 2}$.  As we have discussed earlier,
the relevant correlator is protected.  Therefore, we can study it in
free field theory:
\begin{eqnarray}
    :{1 \over N!} \epsilon_{i_{1}\cdots i_{N}} \epsilon^{j_{1}  \cdots j_{N}}
    \Phi^{i_{1}}_{j_{1}} \cdots \Phi^{i_{N}}_{j_{N}}: \,
        \, :\Phi^{\dagger a_{1}}_{a_{2}} \Phi^{\dagger a_{2}}_{a_{1}}:
    &\propto&
    {1 \over (N-2)!}
    \epsilon_{i_{1}\cdots i_{N-2}a_{1}a_{2}} \epsilon^{j_{1}\cdots
    j_{N-2}a_{1}a_{2}}
\Phi^{i_{1}}_{j_{1}} \cdots \Phi^{i_{N-2}}_{j_{N-2}}
\nonumber \\
&\equiv& {\rm det}_{N-2} \Phi
\end{eqnarray}
where we wrote only the relevant piece in the OPE and have suppressed
spatial dependences as usual.  This suggests that it should be the
operator which creates the single giant state of charge $N-2$.  One
can proceed in this way to compute candidates for lower angular
momentum giant gravitons.

Above we wrote the chiral operator in the OPE and matched it to the
giant graviton of R-charge $N-2$ because the latter should be the most
likely outcome of the scattering process in the bulk.  Of course,
there is also some non-zero probability of decaying to an even smaller
giant graviton while emitting the rest of the angular momentum in low
energy fluctuations.  However, since the brane changes its
configuration only slightly in such processes (its radius changes by
about $R/N$), one expects a good description within the effective
theory on the brane.  Examination of the world-volume Born-Infeld
action shows that all processes in which additional gravitons are
emitted are suppressed by powers of $N$. Likewise, corrections to our
analysis suggesting that the $N-2$ giant graviton is the operator on
the RHS of (\ref{sogrv}) are also supressed by powers of $N$.

An additional heuristic argument suggesting the relevance of
subdeterminants is obtained by considering giant gravitons in $SU(N)$
gauge theory Higgsed by a generic Higgs VEV $V$ to $SU(N-k)\times
U(1)^{k-1}$.  The supergravity description of this theory involves a
domain wall in $\ads{5} \times \sph{5}$.  Near infinity, the sphere
has a radius $R_{\infty} = (g_{s} N \alpha^{\prime 2})^{1/4}$, while
near the origin the sphere radius is $R_{0} = (g_{s} (N - k)
\alpha^{\prime 2})^{1/4}$.  A maximal giant graviton near the AdS
boundary will have a size $R_{\infty}$.  However, a maximal giant
graviton near the origin of the space has a radius $R_{0}$ and should
be created by a determinant of the scalar field of the infrared
$SU(N-k)$ gauge theory.  What operator in the UV $SU(N)$ theory
descends to the determinant of $SU(N-k)$?  One such operator is
$\det_{N-k}(\Phi - V)$ where $\Phi$ is a scalar of the $SU(N)$
theory.\footnote{We have not carefully examined whether any gauge
invariant operators that are zero in the infrared can be added to
$\det_{N-k}(\Phi - V)$.} Upon removing the Higgs VEV ($V \rightarrow
0$) giants of the size $R_{0}$ are less than maximal and the argument
above suggests that these states are created by $\det_{N-k}\Phi$.

Inspired by these observations we introduce, as discussed above, a
family of ``subdeterminant'' operators $O_l \equiv \det_{l}\Phi$,
given in (\ref{subdets}), which have a bound on their dimension and
approach the determinant as $l \rightarrow N$.  The operator $O_l$
carries the same quantum numbers as the giant graviton carrying
angular momentum $l$ in the bulk theory.  The $O_{l}$ enjoy a bound on
their dimension that matches the bound on angular momenta of giant
gravitons: it is clear from the definition~(\ref{subdets}) that $l
\leq N$.  When $l=N$ we recover the determinant as the description of
the maximal giant.

\paragraph{$\bf SO(2N)$:}
As another example, let us consider the $\CN=4$, $SO(2N)$ theory.  We
have shown that the maximal giant graviton of the $SO(2N)$ theory is
$\det(\Phi)$, and expect that giants of size close to maximal will be
created by similar operators.  As before, a simple way of deriving a
candidate set of operators is to consider a process where a small
perturbative graviton is absorbed by a maximal giant.  We can study
this in the field theory by identifying the chiral $2N-2$ R-charge
operator in the OPE of $\det\Phi$ and $\Tr \Phi^{\dagger 2}$.  As we
have discussed earlier, the relevant correlator is protected, and can
be studied in free field theory:
\begin{eqnarray}
     :\Phi^{\dagger a_{1}}_{a_{2}}  \Phi^{\dagger a_{2}}_{a_{1}}:
     \ \ \
    :\epsilon_{i_{1}\cdots i_{2N}} \, \Phi_{i_1i_2}\cdots
    \Phi_{i_{2N-1}i_{2N}} \, \,
     \epsilon^{j_{1}\cdots j_{2N}} \, \Phi_{j_1j_2}\cdots
     \Phi_{j_{2N-1}j_{2N}}: \ \
    \propto \ \ \ \ \ \ \ \ \ \ \ \ \ \ \ \ \ \ \ \ {} \nonumber
    \\
    :\epsilon_{abi_{3}\cdots i_{2N}} \,  \Phi_{i_3i_4}\cdots
     \Phi_{i_{2N-1}i_{2N}} \, \,
     \epsilon^{abj_{3}\cdots j_{2N}}\,
        \Phi_{j_3j_4}\cdots \Phi_{j_{2N-1}j_{2N}}:
\label{sogrv}
\end{eqnarray}
The operator on the right side is not the product of any two chiral
operators with charges $l$ and $2N-2-l$.   Therefore it should be the single
giant state of charge $2N-2$.

\subsection{Overlaps}

The subdeterminant $\det_l$ of a diagonal matrix is the symmetric
polynomial of eigenvalues $s_l$ defined in the Appendix~\ref{detexp}.
Therefore, subdeterminants do not interpolate between traces and
determinants as $l$ increases from a small to a large value.  Rather,
they are simply a good description of large giant gravitons; unlike the
single- and multi-trace operators, the operators $\det_\ell \Phi$ for
$\ell\sim N$ create orthogonal states at large $N$.  We will now
give evidence for this claim, by calculating some correlation
functions of subdeterminants $O_l=\det_l\Phi$ in the $\CN = 4$,
$SU(N)$ conformal field theory.

Specifically, we study (i) the overlap of a maximal giant graviton with a
state containing a smaller giant and a pointlike graviton, and (ii)
the overlap of a maximal giant graviton with two smaller giant gravitons
of half its size.  We will compute the $N$ dependence of these processes
in field theory, using the fact that the relevant correlators are
protected as described in the Sec.~\ref{tools}.  Earlier, in
Sec.~\ref{nottrace}, we examined (i) and (ii) with the assumption that
giant gravitons were made by traces and found that the breakdown of
planarity suggested that the overlaps did not vanish at large $N$.
The planar diagrams by themselves suggested that these overlaps were
$O(1)$ and $O(\sqrt{N})$ (eqs.  (\ref{2pt1}) and
(\ref{2ptgrow})).\footnote{Since breakdown of planarity in the $SU(N)$
theory makes the exact computation of the correlator (\ref{2pt1})
difficult, we examined the case of $U(1)^{N}$ quantum mechanics.  In
this eigenvalue approximation, which we do not present here, the
exact correlation function is $O(1)$.} Fortunately, if giants are made by
subdeterminants, the combinatorics can be done exactly in some cases
(and well-estimated in others) so we will not
need to appeal to the planar approximation.  We will find that the
overlap (i) is $O(1/N)$ while (ii) is exponentially suppressed in
accord with expectation.

If our identification of the operators dual to the giant gravitons is
correct, the overlap of the state created by $O_N$ with the state
created by $O_{N-2} \Tr \Phi^2$ is related to the amplitude for a
maximal giant to split into a giant graviton of angular momentum $N-2$
and a point-like graviton of angular momentum $2$.  We can do the
combinatorics exactly in free field theory. We will need
\begin{eqnarray}
\langle  O_N^{\dagger} \, O_N \rangle &\sim& N! \label{norm1}\\
\langle (O_{N-2}^{\dagger}\Tr \Phi^{\dagger 2}) \, (O_{N-2}\Tr \Phi^2) \rangle
& \sim & 4(N^2+2N+2) N!
\end{eqnarray}
where we have dropped spatial dependences.  The desired overlap
is given by
\begin{equation}
\frac{\langle\, O_N^{\dagger} \, O_{N-2}\Tr\Phi^2 \, \rangle}{\sqrt{\langle\,
O_N^{\dagger} \,
O_N \, \rangle}\sqrt{ \langle\,( O_{N-2}^{\dagger} \Tr \Phi^{\dagger 2}) \,( O_{N-2}\Tr\Phi^2)
\, \rangle  }}=\frac{1}{\sqrt{N^2+2N+2}} \sim \frac{1}{N}
\label{2ptsup}
\end{equation}
This $1/N$ suppression shows that a maximal brane is orthogonal,
in the large $N$ limit, to a state with a near-maximal brane and a
pointlike graviton; however, the amplitude for a brane to emit
a pointlike graviton is perturbative in $1/N$.
This is analogous to the orthogonality property
(\ref{ortho}) for pointlike gravitons of small
angular momentum.

Next  we calculate the $N$ dependence of the correlation function
$\langle O_N^{\dagger} (O_{N/2} O_{N/2}) \rangle$.   This represents
the overlap between a maximal giant and two others of half-maximal
size, and gives the amplitude for a giant to split into two giants of
half the angular momentum.  Since maximal and half-maximal giants are
physically located far from each other on the $\sph{5}$, this overlap
should be heavily suppressed.  We need to compute
\begin{equation}
\frac{\langle O_N^\dagger \, (O_{N/2} O_{N/2}) \rangle}
{\sqrt{\langle O_N^\dagger \, O_N \rangle}
{\sqrt{\langle  (O_{N/2} O_{N/2})^\dagger\, (O_{N/2} O_{N/2}) \rangle}}}
\label{2ptexpa}
\end{equation}
where the denominator is required for proper normalization.
We can compute this in free field theory, since this correlation function
is protected.  First,
\begin{equation}
\langle O_N^{\dagger} \, (O_{N/2} O_{N/2}) \rangle= (\frac{N}{2}!)^2 N!
\end{equation}
This result is exact.  In addition we need Eq.~(\ref{norm1}).
We have not been able to evaluate the last two-point function
exactly, but we have found upper and lower bounds
\begin{equation}
 2(\frac{N}{2}!)^2
N!^2<\langle  (O_{N/2} O_{N/2})^\dagger\, (O_{N/2} O_{N/2}) \rangle <
N!^3
\label{sdet2sdet2}
\end{equation}
as discussed in Appendix~\ref{bounds}.  The normalized
two point function is thus\footnote{It is also easy to compute the
normalized three point function
\begin{equation}
{\langle O_{N}^{\dagger} \, O_{N/2} \, O_{N/2} \rangle
\over
\sqrt{O_{N}^{\dagger} \, O_{N}}
\,
\sqrt{O_{N/2}^{\dagger}\,  O_{N/2}}
\sqrt{O_{N/2}^{\dagger}\,  O_{N/2}}} \sim
{1 \over 2^{N}} \, .
\end{equation}}
\begin{equation}
\frac{1}{2^N} <\frac{\langle O_N^\dagger \, (O_{N/2} O_{N/2}) \rangle}
{\sqrt{\langle O_N^\dagger \, O_N \rangle}
{\sqrt{\langle  (O_{N/2} O_{N/2})^\dagger\, (O_{N/2} O_{N/2}) \rangle}}}
<\frac{1}
{2^{N/2}} \label{2ptexp}
\end{equation}
This amplitude, related to the amplitude for a giant graviton of angular momentum $N$
to split into two giant gravitons of angular momentum $N/2$, is
indeed exponentially suppressed, as we would have expected.

The suppression of the two-point functions (\ref{2ptsup}) and
(\ref{2ptexp}) is in marked contrast to (\ref{2pt1}) and
(\ref{2ptgrow}), showing that, as a basis of high-dimension operators
at large $N$, subdeterminants are certainly an improvement over
traces. It would be very interesting to compute the bulk transition
amplitudes between states with different configurations of giants, and
to compare them with field theory predictions.

\newcommand{\theory}[0]{(2,0)_{k}}
\section{Giants on $\ads{7}\times\sph{4}$ -- a Matrix model analysis}
\label{matrix}

We have presented topological arguments and free field computations of
correlation functions that strongly suggest that giant gravitons on
$\ads{5} \times \sph{5}$ and its orbifolds are dual to states created
by subdeterminants.  In this section, we will study giant gravitons on
$\ads{7} \times \sph{4}$, which is dual to the $\theory$ CFT
describing the low energy dynamics of $k$ coincident 
M5-branes~\cite{mald}.  It is
believed\footnote{A matrix theory argument for this is given
in~\cite{onm}.  The AdS/CFT correspondence leads to similar
conclusions for low momenta~\cite{ofya}.  There is also a direct field
theory argument from the reduction of the $(2,0)$ theory on a circle
to $4+1$-dimensional Yang-Mills theory.} that the chiral primaries of
this theory, as in $d=4, \, \CN =4$ theory are scalar operators in the
symmetric, traceless representations of the R-symmetry $SO(5)$. On the
gravity side, the $\ads{7} \times \sph{4}$ background has giant
graviton excitations \cite{suss} which are M2-branes wrapping an
$\sph{2}$ within $\sph{4}$ and moving with some angular momentum,
matching the charges of the $\theory$ chiral primaries. In addition to
the giant gravitons being M2-branes, the other difference from the
$\ads{5}\times \sph{5}$ case is that the giant graviton radius/angular
momentum relation is different and it is now $r=R_sl/N$, where $R_s$
is the radius of the $S^4$.

We do not have a Lagrangian description of the $\theory$, and so we
cannot explicitly construct operators from elementary fields.
Nevertheless, we will show that the expectation value in the Coulomb
branch of the operator creating a nearly maximal giant is a
subdeterminant of a diagonal matrix made out of the Higgs VEVs.
Because of the relation between the $(2,0)$ theory and $\CN=4$
Yang-Mills in four dimensions, this is analogous to evaluating the
expectation value of a near-maximal giant graviton in the Higgsed
$\CN=4$ theory, which we previously argued is a subdeterminant.

We will pursue the following strategy:\footnote{Parts of
this section were developed in collaboration with Moshe Rozali.}
\begin{itemize}
    \item We recall the Discrete Light Cone Quantization (DLCQ) of the
    $\theory$ theory as a Matrix model, and its relation to the
    lightcone quantization of the Poincar\'e patch of
    $\ads{7} \times \sph{4}$.
    \item We identify the giant graviton states in this Matrix model,
    relying on a desccription of the $\sph{4}$ via  non-commutative geometry.
   \item Using the DLCQ description, we compute the expectation value
    of a chiral giant graviton with R-charge $k-2$ along the Coulomb
    branch, and find that it is a subdeterminant of the VEVs.
\end{itemize}
This further justifies the identification of giant gravitons with
subdeterminants.

We will carry out our DLCQ calculations with one unit of lightcone
momentum.  Strictly speaking, one needs to take the limit $({\it null
\ momenta}) \rightarrow\infty$ in order to reliably obtain the
uncompactified $\theory$ theory.  Nevertheless, since we focus on
correlators which are presumably protected by SUSY\footnote{Their
$\CN=4$, $D=4$ counterparts are protected.}, we expect that our
computation will yield the correct physical result, as is often the
case in Matrix theory.

\subsection{Review of the model}

\paragraph{The model}

The M(atrix)/DLCQ \cite{bfss,lnydlcq} description of the $\theory$
theory with $N_{0}$ units of momentum along the null circle is quantum
mechanics on the moduli space of $N_{0}$ instantons in four
dimensional $\CN=4,\ U(k)$ gauge theory \cite{mcmike,esnom,onm}.
The instanton moduli space is a finite dimensional hyperkh\"{a}ler
quotient given by the ADHM construction.  We will focus on $N_0=1$,
which is particularly simple.  In this case the moduli space splits
into an $R^4$ component, which parametrizes the instanton's center of
mass, and a hyperk\"{a}hler component of real dimenson $4k-4$ which
encodes the size and orientation of the instanton.

The latter manifold is constructed as follows.
Take two fields $Q$ and $\tilde{Q}$ transforming as  ${\bf (1,k)}$
${\bf (-1,\bar{k})}$ under
${U(N_0=1)}_{{\rm gauge}}\times {U(k)}_{{\rm global}}$.   The single
instanton moduli space is then described by solutions to
\begin{equation}
\biggl\{
Q_i{Q^i}^*-{{\tilde Q}_{i*}} {\tilde Q}^i=0,\ \
Q_i{\tilde Q}^i=0
\biggr\}/U(1) \, ,
\end{equation}
where the lower (upper) $i$ index is a fundamental (anti-fundamental)
of $U(k)$.  This is the Higgs branch of a $D=4,\ {\cal N}=2$ field
theory with the same field content.  The bosons also have fermionic
superpartners which we will denote by $\mu^A_i$ and their conjugate
${\bar\mu}_{A}^{i}$, where $A$ is an index in the ${\bf 4}$
representation of an $SO(5)\sim Sp(4)$ R-symmetry - in the DLCQ
interpretation, the R-symmetry is identified with the rotations
transverse to the M5-branes.  The $\mu$'s ($\bar\mu$) also have charge
$1$ ($-1$) under the gauge symmetry.

In lightcone coordinates, the $\theory$ theory has a spectrum of
chiral operators $O_l(x^+,x^-,x^{1..4}_\perp)$ which act on the vacuum
$\vert \Phi \rangle_{{\rm FT}}$ to create states.\footnote{Of course,
to make these states normalizable, we should smear them in the
standard fashion.} In particular \cite{onm}, we can construct states
with finite, positive lightcone momentum $P_{-}$, which will remain
visible in the DLCQ quantum mechanics.  In other words, there should
be chiral states on the ADHM moduli space, with a spectrum determined
by the chiral operators in the $(2,0)_k$ field theory.

Indeed, there is a cohomology problem on the ADHM moduli space
\cite{onm} (for arbitrary $k$ and instanton number) that counts such
chiral states.  It transpires that there is one generator of the
chiral ring for each symmetric, traceless tensor representation of
$SO(5)$ with $1$ up to $k$ indices.  The $1$ representation is a free
tensor multiplet that is related to the center mass of the 5-branes
and will be ignored by us.  The remaining $k-1$ representations are in
the interacting conformal field theory.

The representatives of the cohomology classes related to the chiral
operators can be chosen to be invariant under the global $U(k)$. Hence
we will restrict our attention to such $U(k)$ invariant states in
general. To specify a $U(k)$ invariant form it suffices to specify it
at one point on each orbit of the $U(k)$ symmetry. The generic orbits
are of co-dimension 1 in the ADHM manifold, and we will parametrize
these orbits by picking the following point on each one:
\begin{equation}
Q_i = (Q,0,0..) ~~~~~;~~~~~
{\tilde Q}^i=(0,Q,0,..)
\label{adhmgauge}
\end{equation}
where $Q$ is real and non-negative.  We parametrize the fermion
superpartners ($\mu_{i}^{A}$ and their conjugates) of $Q,{\tilde Q}$ as
\begin{equation}
    (\eta_{1}^{A},\eta_{2}^{A}, \nu_{1}^{A} \cdots \nu_{k-2}^{A})
    \label{musplit}
\end{equation}
and their conjugates.  Some of the $\eta$'s are set to zero by the
fermionic superpartners of the ADHM constraints.

The $U(k)$ invariant wave functions, in this gauge, are given by an
arbitrary function of $Q$ times a state in the Fock space generated by
($\eta, {\bar\eta}$), times a $U(k-2)$ invariant state in the
($\nu, {\bar\nu}$) Fock space. In addition we need to require that the
state is $U(1)_{gauge}$ invariant. Let us call such a function
\begin{equation}
{|\Psi(Q,\eta,{\bar\eta},\nu,{\bar\nu})\rangle}
 = f(Q) \, {| \Psi\rangle}_{\eta\bar\eta} \, {|\Psi\rangle}_{\nu\bar\nu}
\label{ggfxd}
\end{equation}

It will be useful for us to exhibit the $U(k)$ invariant wave function
in another way. Let us view (\ref{ggfxd}) as specifying the
wave function on the submanifold (\ref{adhmgauge}). Since this
submanifold intersects each generic orbit of $U(k)$ once, it lifts to
a unique $U(k)$ invariant function on the ADHM moduli space, obtained
by averaging over the action of the group on this function (this
action moves the supprt of the function from (\ref{adhmgauge}) to other
places in the manifold). We will write this as:
\begin{equation}
{|\Psi\rangle}_{inv}= \int_{U(k)} dg \, 
\bigl(g*{|\Psi(Q,\eta,{\bar\eta},\nu,{\bar\nu})\rangle}\bigr)
\label{invfnc}
\end{equation}

\medskip

\paragraph{Mapping to $AdS_7\times S^4$:}
To go to the description in terms of a light cone version of
$AdS_7\times S^4$, one then defines the following 5 collective
coordinates \cite{doreyandco,michaherman,michaofer}
\begin{equation}
Y^{[AB]}={
{\nu^A_k{\bar\nu}^k_C J^{BC}-\nu^B_k{\bar\nu}^k_C J^{AC}
+{1\over2}J^{AB}{\nu^F_k}{\bar\nu}^k_F +\
\eta\bar\eta \ {\rm terms}}
\over
Q^2} \, .
\end{equation}
Here $J^{AB}$ is the anti-symmetric form of $Sp(4) \sim SO(5)$, which
we take to be $J^{13} = - J^{31} = J^{24} = - J^{42} = 1$.  The
details of the $\eta\bar\eta$ terms are suppressed as they will not
play an important role in the computations below.  $A$ and $B$ are
indices in the ${\bf 4}$ of $SO(5)$ and $[AB]$ denotes the ${\bf 5}$
of $SO(5)$.

These 5 collective variables, in addition to the 4 instanton center of
mass coordinates, in addition to a pair of lightcone coordinates,
describe the position of a particle in a lightcone description of the
Poincar\'e patch of $\ads{7} \times \sph{4}$.  The lightcone dynamics
of $\ads{7} \times \sph{4}$ is precisely described by quantum mechanics on
the ADHM moduli space, when written in terms of the above collective
coordinates \cite{michaherman}.

For the purposes of identifying the various giant gravitons, it is
convenient to isolate the coordinates of the sphere out of the ADHM
collective variables described above.  The space of functions on the
sphere is constructed out of the Fock space of the fermion $\nu$, and
realizes a fuzzy 4-sphere \cite{michaherman}.  We therefore split up
the five $Y$ coordinates as follows.  The radial coordinate of AdS is
identified with
\begin{equation}
U^2={2(k-2)\over R_s Q^2}
\label{ucoor}
\end{equation}
where $R_s=l_p{(\pi k)}^{1\over 3}$ is the radius of the $S^4$,
and the four coordinates of the sphere are
\begin{equation}
{\cal B}^{[AB]}=
   {R_s\over 2(k-2)}
\left( \nu^A_k{\bar\nu}^k_C J^{BC}-\nu^B_k{\bar\nu}^k_C J^{AC}
+{1\over2}J^{AB}{\nu^F_k}{\bar\nu}^k_F \right)
\label{bilinears}
\end{equation}
These ${\cal B}^{[AB]}$ coordinates parametrize a fuzzy $S^4$ (for
details on fuzzy 4-spheres in general and in M(atrix) theory, and their
potential use in the AdS/CFT correspondence, see
\cite{taylor,oldgrv}).

Now we will identify the chiral states of the ADHM quantum mechanics
with states on $\ads{7} \times \sph{4}$.  Let us restrict attention to
the wavefunctions on $\sph{4}$, i.e. to the Fock space of the
bilinears of the $\nu$ fermions.\footnote{This gives the scalar
spherical harmonics on the $\sph{4}$.  To recover the tensor harmonics
one needs to refine the analysis to include the $\eta$ fermions. We
will provide the relevant information on this refinement later.} The
detailed derivation is described in~\cite{michaherman}; here we quote
the result.  Let $\vert\Phi \rangle$ be the state which is annihilated
by all the $\bar\nu$.  Then the constant function on the fuzzy $S^4$
is given by
\begin{equation}
    \vert 1 \rangle = \epsilon^{i_1 \cdots i_{k-2}}\epsilon^{j_1 \cdots j_{k-2}}
(J_{A_1B_1}\nu^{A_1}_{i_1}\nu^{B_1}_{j_1}) \cdots
(J_{A_{k-2}B_{k-2}}\nu^{A_{k-2}}_{i_{k-2}}\nu^{B_{k-2}}_{j_{k-2}})
\vert\Phi\rangle
\label{one}
\, .
\end{equation}
It may be surprising that $|\Phi\rangle$ is not itself the constant
function on the sphere.  This is so because $|\Phi\rangle$ is not
invariant under the $U(1)_{{\rm gauge}}$; the $2k-2$ fermions $\nu$ in
(\ref{one}) lead to a net charge zero for the state $|1\rangle$ as
required.

To construct higher spherical harmonics we act  with the  bilinears
$\CB$ on the vacuum $\vert 1\rangle$,  after symmetrizing
the product by contraction with a symmetric traceless tensor $T_{i_1
\cdots i_l}$:
\begin{eqnarray}
\vert T_{[A_1B_1] \cdots [A_lB_l]} \rangle &=& T_{[A_1B_1] \cdots [A_lB_l]}
{\cal B}^{[A_1B_1]} \cdots {\cal B}^{[A_lB_l]}\vert 1 \rangle
\nonumber \\
&\propto&
 T_{s_1 \cdots s_l}\epsilon^{i_1\cdots i_{k-2}}\epsilon^{j_1 \cdots j_{k-2}}
(\Gamma^{s_1}_{A_1B_1}\nu^{A_1}_{i_1}        \nu^{B_1}_{j_1}) \cdots
(\Gamma^{s_l}_{A_lB_l}\nu^{A_l}_{i_l}        \nu^{B_l}_{j_l})
\label{states}
\\
&& \ \ \ \ \ \ \ \ \ \ \ \ \ \ (J_{A_{l+1}B_{l+1}}   \nu^{A_{l+1}}_{i_{l+1}}\nu^{B_{l+1}}_{j_{l+1}})
\cdots
(J_{A_{k-2}B_{k-2}}   \nu^{A_{k-2}}_{i_{k-2}}\nu^{B_{k-2}}_{j_{k-2}})
\vert\Phi\rangle \, .
\nonumber
\end{eqnarray}

\paragraph{Algebraic Aspects:}
Defining the $SO(5)$ generators on the fermions $\nu$,
$$M^{\{AB\}}={1\over2}\nu^A_l{\bar\nu}_C^lJ^{BC}+
         {1\over2}\nu^B_l{\bar\nu}_C^lJ^{AC},$$
where $\{AB\}$ denotes symmetrization on the indices, we obtain that
the coordinates of the fuzzy 4-satisfy the algebra
\begin{equation}
[{\cal B}^{[AB]},{\cal B}^{[CD]}]\propto J^{AC}M^{\{BD\}}-
                    J^{BC}M^{\{AD\}}-
                    J^{AD}M^{\{BC\}}+
                    J^{BD}M^{\{AC\}}
\label{alg1}
\end{equation}
Given that ${\cal B}$ are a ${\bf 5}$ of $SO(5)$, we can define
$M^{6i}={\cal B}^i,\ i=1..5$ to obtain an $SO(6)$ algebra.

The Casimir of this $SO(6)$ algebra is proportional to
\begin{equation}
J_{AC}J_{BD}{\cal B}^{[AB]}{\cal B}^{[CD]}-
{R_s^2\over 4(k-2)^2}J_{AC}J_{BD}M^{\{AB\}}M^{\{CD\}}
\end{equation}
(Note that we have chosen the normalization factor in accordance
with (\ref{bilinears}).)
If we denote by ${\hat N}$ the fermion number operator $\nu^A_l{\bar\nu}^l_A$,
one can evaluate this expression to be
\begin{equation}
{R_s^2\over 4(k-2)^2}
\biggl(4(k-2){\hat N}-{\hat N}^2+4{\hat N}-4C_{2,U(k-2)}\biggr)
\end{equation}
where the last term is the casimir of $U(k-2)$: 
$\ C_{2,U(k-2)} = (\nu^A_l{\bar\nu}^s_A)(\nu^A_s{\bar\nu}^l_B)$.

Evaluating the expression on the states (\ref{states}) we obtain, at
large $k$, that
\begin{equation}
\biggl(
J_{AC}J_{BD}{\cal B}^{[AB]}{\cal B}^{[CD]}
- {R_s^2\over 4 k^2}J_{AC}J_{BD}M^{\{AB\}}M^{\{CD\}}\biggr)
| T_{[A_1B_1] \cdots [A_{k-2}B_{k-2}]}\rangle
=R_s^2| T_{[A_1B_1] \cdots [A_{k-2}B_{k-2}]}\rangle
\label{realcas}
\end{equation}

\subsection{Giant gravitons}

In this section we will identify the states (\ref{states}) as giant
gravitons on $\sph{4}$, which are membranes wrapping $\sph{2}$s inside
$\sph{4}$.  Consider the $S^4$ as embedded inside $R^5$ (with
coordinates $\{X^{1} \cdots X^{5}\}$) in the obvious way, and assume
that the giant graviton carries angular momentum $l$ in the
$X^{1}-X^{2}$ plane.  This object blows up into an M2-brane wrapping a
2-sphere in $X^{3,4,5}$~\cite{suss}.  The radius of the membrane is
$r^{2} = {X^3}^2+{X^4}^2+{X^5}^2={L^2\over N^2}R_s^2,$ where $R_s$ is
the radius of $\sph{4}$.  Projected onto the $X^1-X^2$ plane, the
M2-brane circles at ${X^1}^2+{X^2}^2=(1-{L^2\over N^2})R_s^2$ and it
is clear that the maximal angular momentum allowed is $L=N$.

\medskip

\paragraph{\bf The Fuzzy 2-Sphere:}
To relate the chiral states of the quantum mechanics to these giant
gravitons, it is helpful to recall the familiar description of the
fuzzy 2-sphere as the geometry seen by dipoles on an $S^2$ in a
magnetic field passing through the sphere. The dipole approach is
developed in \cite{suss}, but we will be more interested in the
algebraic aspects. We will develop these in detail and then reason by
analogy about spherical membranes on $\sph{4}$. Some related work for
the fuzzy $S^2$ appears in \cite{nair}.

The fuzzy 2-sphere is realized by the matrix algebra
\begin{eqnarray}
[W^i,W^j]&=&i \epsilon^{ijk}W^k, \ \ i=1\cdots 3 \nonumber \\
{W^i}^2&=& {N\over 2}({N\over2}+1) \nonumber
\end{eqnarray}
The spherical harmonics on this fuzzy sphere are constructed from the
$W^{i}$ as symmetric traceless polynomials with $0$ to $N$
indices.

To interpret these harmonics as configurations of dipoles in a
magnetic field, it is helpful to examine the flat space limit that
arises by working around the north pole ($W^{3}$ maximal) and taking
$N \rightarrow \infty$.  Then
\begin{equation}
[W^1,W^2]= {\rm Const. }\ + \ O((W^{1})^{2} + (W^{2})^{2}) \ {\rm corrections}
\, .
\end{equation}
This is the (approximate) algebra of the fuzzy $R^{2}$.  In this case,
it is known that the coordinates $W$ parametrize the position of one
of the tips, say the positive charge end, of a dipole moving in a
constant magnetic field \cite{bigsuss,sbrgwtn}.  This is also the case
for the dipole on the fuzzy $S^{2}$.  Of course, we could just as well
describe the other tip of the dipole on the fuzzy $\sph{2}$ which
satisfies the algebra
\begin{equation}
[{\hat W}^i, {\hat W}^j]=-i \epsilon^{ijk}{\hat W}^k \, .
\end{equation}

We wish to generalize these relations to the case of spherical
membranes whose dynamics will sense a fuzzy $\sph{4}$.  Unlike the
dipole, there is no preferred endpoint on a spherical membrane.  So we will
parametrize the membrane in terms of its center and radius.   If we
regard the $\sph{4}$ as embedded in $R^{5}$, the membrane center that
we will use is the center of mass in $R^{5}$.  Hence it can lie off
the $\sph{4}$, although all points on the membrane are of course
restricted to this sphere.  It is helpful to examine how such a
parametrization affects the dipole description of a fuzzy $\sph{2}$.

In the latter case, define the center of mass coordinate and the size
of the dipole as
\begin{eqnarray}
W_{cms}^i &=& {1\over 2} (W^i+{\hat W}^i) \nonumber \\
\hat{M}^{jk}=\epsilon^{ijk}\Delta^i &=& \epsilon^{ijk}{1\over 2}
(W^i-{\hat W}^i)
\, . \nonumber
\end{eqnarray}
Here we regard the $\sph{2}$ as embedded in $R^{3}$, and permit the
center of mass and length of the dipole to take values in $R^3$ and
not just on $\sph{2}$.  Then $W_{cms}^i$ and $\hat{M}^{ij}$ satisfy the
following algebraic relations
\begin{enumerate}
\item $\hat{M}^{ij}$ forms an $SO(3)$ algebra.
\item $W_{cms}^i$ is a ${\bf 3}$ of this $SO(3)$.
\item $[W_{cms}^i,W_{cms}^j]\propto \bar{M}^{ij}$
\end{enumerate}
These relations describe the motion of dipoles on a fuzzy $\sph{2}$.
They are identical, after replacing $SO(3)$ by $SO(5)$, to the algebraic
relations governing the motion of giant gravitons on $\sph{4}$ (see
(\ref{alg1}) and the surrounding discussion).  Note that $W_{cms}$
and $\hat{M}$ together form an $SO(4)$ algebra.

Spherical harmonics on the fuzzy $\sph{2}$ are symmetric traceless
representations of $SO(3)$ with $0$ up to $N$ indices.  When the
symmetry is promoted to $SO(4)$, all of these representations form a
single symmetric traceless $SO(4)$ representation with $N$ indices.
Hence the $SO(4)$ Casimir, which we will denote by $C_2$, is constant
on all the states of the fuzzy $\sph{2}$.  That is,
\begin{equation}
C_2 = ({W_{cms}^i})^2+{(\hat{M}^{ij})}^2 =
{(W^i_{cms}\pm\Delta^i)}^2 
\label{dipoleeq}
\end{equation}
The first equality is the Casimir of $SO(4)$ and the second follows
from our definitions of $\hat{M}$ and $W$.  We have also used the fact
that $\sum_{i} W^{i}\, \Delta^{i} = 0$.  The fact that $C_{2}$ is
fixed on all the states of the fuzzy sphere (since they form a single
$SO(4)$ representation) has a clear physical interpretation: the tips
of the dipole are on $\sph{2}$.  Below we will a similar structure for
the fuzzy 4-sphere and giant gravitons on $\sph{4}$.

\medskip

\paragraph{\bf The fuzzy 4-Sphere:} The structure that we have obtained
for the fuzzy $\sph{2}$ is remarkably similar to the structure
uncovered in the previous section of the fuzzy $\sph{4}$ description,
via DLCQ quatum mechanics, of the sphere in $\ads{7} \times \sph{4}$.
We are therefore led to the following identifications:
\begin{enumerate}
\item The $M^{[ij]}$ in the $\sph{4}$ algebra
are analogous to the $\hat{M}^{[ij]}$ appearing in the dipole
description of $\sph{2}$.  In analogy, they parametrize the size of
the spherical M2-brane. Indeed, the radius of the giant graviton satisfies
$r^2\propto -{M^{[ij]}}^2$ \cite{suss}.
\item We identify $\CB^{i}$ as the center of mass
coordinates of a spherical M2-brane moving on $\sph{4}$.
\end{enumerate}
In both these expression we switched to an $SO(5)$ vector notation
$\CB^i=\Gamma^i_{[AB]}\CB^{[AB]}$.

Finally, the Casimir condition (\ref{realcas}) is identical to the
requirement that the membrane is on the $S^4$. Under the
identification of the size and center of mass position of the
membrane, requiring that the membrane is on the sphere implies
$${\CB^i}^2-{R^s\over k^2} {M^{[ij]}}^2=R_s^2$$ for all states on the
fuzzy 4-sphere (in our conventions ${M^{ij}}^2$ is negative definite),
which is the same as (\ref{realcas}).

Hence all the chiral states in the theory (i.e., (\ref{states})) are
naturally interpreted as giant gravitons on the $\sph{4}$ factor of
$\ads{7} \times \sph{4}$, according to the interpretation of $\CB$ as
the center of mass coordinates and the Casimir equation as the
requirement that the M2-brane is on the 4-sphere. In fact, all the
dynamical information regarding the giant graviton is automatic if one
identifies the $\CB$'s in this way, and uses the Casimir equation.

\subsection{Computation of the expectation value}

Having identified the chiral states of the ADHM construction as giant
gravitons, we can compute the expectation value of a giant graviton
operator along the Coulomb branch of the $\theory$ theory, along which
the 5-branes are at separated positions.  We will find that this
expectation value is a subdeterminant of the 5-brane position
eigenvalues, rather than a Trace of these quantities.

The computation will be almost trivial once we set up a convenient
kinematical framework. To do so we need to explain:
\begin{enumerate}
    \item The details of the configuration along the Coulomb branch, and its
   Matrix realization.
    \item The precise expectation value that we are computing
\end{enumerate}

\paragraph{\bf Flat directions of the $\theory$ and
their Matrix model realization:} The moduli space of the $\theory$
theory is parametrized by the positions of the $k$ 5-vectors, each in
the {\bf 5} of the $SO(5)$ R-symmetry. Since these are identical
branes, the moduli space is actually $R^{5k}/S_k$.  It is
straightforward to write down the quantum mechanics which describes
the DLCQ of the Higgsed theory.  The original quantum mechanics was
the Higgs branch of the D0-D4 system, with $1$ D0-brane and $k$
D4-branes.  Higgsing the 5-brane theory corresponds to separating the
positions of the D4-branes, i.e, to changing the expectation values of
the 4-4 strings which parametrize the transverse position of the
D4-branes. Instead of being 0, these 5 $k\times k$ matrices are taken
to be arbitrary commuting matrices. This in turn, changes the D0-brane
quantum mechanics by the introduction of background value for scalars
in the vector multiplets of the $U(k)$ flavor symmetry.

The terms which are relevant for us are the ones that include the
fermions ($\nu, \bar\nu$) and ($\eta, \bar\eta$).  Denoting these 5
commuting matrices by ${\lambda^{[AB]}}^i_j$, where $[AB]$ is an index
in the ${\bf 5}$ of $SO(5)$, and $i,j$ are $U(k)$ indices, these
couplings are:
\begin{equation}
\delta H=\Sigma_{i,j=1}^k \, \, {\lambda^{[A'B']}}^i_j \, \, J_{A'A}
\, \, J_{B'B} \, \, \mu^A_i{\bar\mu}^{Bj}
\label{couplings}
\end{equation}
where for notational convenience we combined $\nu$ and $\eta$ fermions
back into the $\mu$ fermions.  

The simplest Higgsing for our purposes is the following.  Consider two
coordinates out of the {\bf 5} of $SO(5)$ and assemble them into a
complex scalar $Z$.  This breaks the $SO(5)$ into $SO(3)\times SO(2)$. 
We will Higgs the theory by giving non-zero expectation values only to
$Z_i=\lambda_i,\ i=1 \cdots k$, i.e, the expectation values of $Z$ and
$Z^{*}$ are charged as $1_1$ and $1_{-1}$ respectively under the
$SO(3)\times SO(2)$.  The fermions $\nu$ are in a ${\bf 4}$ of $SO(5)$
which splits into a ${\bf 2}_{1\over2}+{\bf 2}_{-{1\over2}}$ under
$SO(3)\times SO(2)$.  We are interested in a bi-linear combination
which transforms as a ${\bf 1}_1$ under this symmetry, determining the
$[A'B']$ pairs that actually appear in (\ref{couplings}).  If we
denote the fermions which are in the ${\bf 2}_{1\over2}$ as $\mu^1_i$
and $\mu^2_i$ then (recalling our conventions for $J_{AA'}$) the
interaction is now $\delta H+(\delta H)^\dagger$ where
\begin{equation}
\delta H = \Sigma_{l,t=1}^k \,
{\bigl(\lambda^{[12]}\bigr)}^l_t \,
(\mu^3_l \, {\bar\mu}^{4t}- \mu^4_l{\bar\mu}^{3t})
\label{fininter}
\end{equation}

\paragraph{\bf The expectation value we wish to calculate:}
To discuss the details of the relevant expectation value, it will be
useful to first specify this computation on $AdS_7\times S^4$, and
then translate it into light cone quantization, and to the DLCQ.

We will also focus on  specific chiral operators. Prior to Higgsing
the $\theory$, the states of the graviton (\ref{states}) (pointlike
for low and giant for high angular momenta) on $S^4$ are in symmetric
traceless representations of $SO(5)$ with $2$ to $k$ indices. Let us
focus on the operator in each of these representations whose quantum
numbers are the same as $Z^l,\ l=2..k$.  We will denote these
operators by $O_l$. Although the Higgs VEVs, which separate the
5-branes, break $SO(5)$ in the infrared, we can still classify the
relevant operators using $SO(5)$ quantum numbers in the ultraviolet.
Equivalently, the Higgs VEVs are described in gravity as domain walls
in AdS that have specific locations on $\sph{4}$, but near the
spacetime boundary all the symmetries of $\ads{7} \times \sph{4}$ are
recovered.

We would like eventually to compute a 1-pt function in the DLCQ model,
but we can do so only indirectly.  It is easy enough to identify
states created by chiral operators in the DLCQ; i.e., states of the
form $O_l| \Phi\rangle_{FT}$, where $| \Phi\rangle_{FT}$ is the
$\theory$ vacuum.  But we can not compute one-point functions of these
operators because in the DLCQ construction they are smeared to carry
non-zero lightcone momentum, whereas the theory has translation
invariance in that direction; hence they cannot have a one-point
function.  In order to compute the expectation value of a zero
lightcone momentum operator we should compute the 2-pt function of
operators with opposite momentum.  We can understand this as an
overlap calculation of an incoming and an outgoing state with the same
momentum.  The operators in question can fuse to give a zero momentum
operator with a one-point function.

A convenient two-point function to compute is $\langle O_2(x)^\dagger
O_k(0) \rangle$.  (We start with the local operators on $R^{5,1}$, and
will later discuss the smearing that gives these objects the lightcone
momentum necessary for a DLCQ analysis.)  We choose this two-point
function because the operators have a maximal difference in R-charges,
and therefore, as we will see, the dependence of the 2-point function
on the VEVs will be clearer.  The two-point function will receive
contributions from all operators in the OPE of $O_{2}^{\dagger}$ and
$O_{k}$.  We will pay attention to the contribution from the chiral
primary operator with R-charge $k-2$, which we will denote by
$O_{k-2}$.  We can extract the contribution of this operator by its
quantum numbers, or using the following alternative method.  This
operator appears in the OPE with the most singular $x$ dependence. 
Hence we can focus on it by taking $x\rightarrow 0$ and rescaling by
${|x|}^4$ (the power is determined by conformal invariance).

To compute the expectation value $\lim_{x\rightarrow 0}\langle
O_2(x)O_k(0) \rangle$ using $\ads{7}\times \sph{4}$ one would insert
two sources on the AdS boundary for the corresponding particles (one
of them being a giant graviton).  In the $x\rightarrow 0$ limit, the
AdS calculation is dominated by a 3-point interaction in the bulk that
occurs near the AdS boundary.  In effect, the fields created by the
sources on the boundary meet and fuse into a third field near the
boundary.  The two-point function is therefore related to the
one-point function of the third field.  As described above, by using
quantum numbers, or by isolating the leading singularity we can
extract from this the 1-point function of $O_{k-2}$

If $O_k$ is the operator that corresponds to the maximal giant
graviton, and $O_2$ the lowest momentum pointlike graviton, then by the
same argument as in Sec.~4, we know that the operator $O_{k-2}$ that we
obtain in the OPE corresponds to the giant graviton of R-charge
$k-2$. This means that we will be computing the expectation value of a
giant graviton operator along the Coulomb branch of the $\theory$ 
theory.

We will work in lightcone coordinates; so, instead of $O(x)$ we use
$O(x^+,P_-,x^{1\cdots 4}_\perp)$, and we will also specialize to the
case $x_\perp^{1..4}=0$. These operators are smeared out on $x^{-}$,
and so in position space we are inserting operators at all values of
$x^{-}$.  Nevertheless, if the difference in the $x^{+}$ position of
two such operators goes to zero, then in position space the distances
between all operators in the computation becomes null, and we can use
the short distance properties of OPE's.  The considerations of the
previous paragraphs then apply again.

We will assume that this carries over to the DLCQ description.  We
have already described states (\ref{states}) that correspond to
$O_l(P_-=1/R)$ acting on the vacuum: $| O_l \rangle$.  Hence, in order
to obtain the 2-pt function $\langle O_2^\dagger O_k \rangle$, we will
compute the overlap between two such states $\langle
O_2(x^+=\epsilon)| O_k(x^+=0) \rangle$.

Let us elaborate on the structure of the wavefunctions $| O_2 \rangle$
and $|O_k \rangle$.  As in (\ref{ggfxd}) and (\ref{invfnc}), we need
to specify 3 parts of each of these functions - i.e, the dependence on
$Q$, and the contributions from the Fock spaces of ($\eta,
{\bar\eta}$) and ($\nu,{\bar\nu}$):
\begin{enumerate}
\item Since we want our states to be localized at the boundary of $AdS$
in the gravity description,
we need to take wave functions with a support at small $Q$. (See
(\ref{ucoor}).)
\item The $SO(5)$ quantum numbers will receive contributions from both
the $\eta$ and the $\nu$ parts of the wave function. To get the
maximal R-charge the $\nu$ contribution must be maximal for $| O_k
\rangle$, and likewise minimal for $| O_2\rangle$, i.e,
\begin{equation}
{\vert O_2 \rangle}_{\nu\bar\nu} =\epsilon^{l_1 \cdots l_{k-2}} \,
\epsilon^{t_1 \cdots t_{k-2}} \, (J_{A_1B_1}\, \nu^{A_1}_{l_1} \,
\nu^{B_1}_{t_1}) \cdots (J_{A_{k-2}B_{k-2}} \, \nu^{A_{k-2}}_{l_{k-2}}
\, \nu^{B_{k-2}}_{t_{k-2}}) \, \vert\Phi\rangle
\end{equation}
and
\begin{equation}
{\vert O_k\rangle}_{\nu\bar\nu} =\epsilon^{l_1..l_{k-2}} \,
\epsilon^{t_1 \cdots t_{k-2}} \, (\Gamma^z_{A_1B_1}\, \nu^{A_1}_{l_1}
\, \nu^{B_1}_{t_1}) \cdots (\Gamma^z_{A_{k-2}B_{k-2}}\,
\nu^{A_{k-2}}_{l_{k-2}} \,\nu^{B_{k-2}}_{t_{k-2}}) \, \vert\Phi\rangle
\label{okstat}
\end{equation}
where $\Gamma=\Gamma^1+i\Gamma^2$ and $Z=X^1+iX^2$ was defined
before. As before (see the discussion around (\ref{fininter})) it is
easy to obtain from the quantum numbers of $O_k$, that the $[A_iB_i]$
pairs should all be pairs of the form $[12]$ out of the {\bf 4} of 
$SO(5)$.
\item The ($\nu,\bar\nu$) contribution to the R-charge is a singlet for
$|O_2\rangle$ and $k-2$ for $|O_k\rangle$. Hence the ($\eta,\bar\eta$)
Fock space state should contribute 2 units of R-change. The details of
this wave function, which we will denote by
${|O_2\rangle}_{\eta\bar\eta}$ will not matter for us, except that it
is the same wave function for both the states.
\end{enumerate}
For our purposes it will be more convenient to use the explicitly
$U(k)$ invariant form of the states in equation (\ref{invfnc}).  I.e.,
we would like to compute the overlap
\begin{equation}
\int_g \int_{g'} dg \, dg' \,
 \biggl(
{}_{\nu\bar\nu} \langle O_2| \times {}_{\eta\bar\eta}\langle O_2|
\times f^*(Q)
 \biggr)_{t=\epsilon}{g'}^\dagger  \, || \, 
 g \biggl(
 f(Q) \times {|O_2\rangle}_{\eta\bar\eta} \times {|O_k\rangle}_{\nu\bar\nu}
\biggr)_{t=0}
\end{equation}
where we have introduced the notation $||$ to indicate an overlap
between the incoming and outgoing states, each contructed from three
components in three separate Hilbert spaces.  We will need to do
computation in the Higgsed $\theory$.

\medskip

\paragraph{\bf The computation}
We can now proceed to compute the expectation value.  We will treat
the interaction (\ref{couplings}) in perturbation theory, and hence we
would like to evaluate expressions like
\begin{equation}
\int_g \int_{g'} dg \, dg' \,
\biggl( {}_{\nu\bar\nu}\langle O_2|\times
 {}_{\eta\bar\eta}\langle O_2| \times f^*(Q) \biggl )_{t=0}{g'}^\dagger
\, || \, (\delta H + \delta H^{\dagger})^n \, || \, 
g \biggl( f(Q) \times {|O_2\rangle}_{\eta\bar\eta}
 \times {|O_k\rangle}_{\nu\bar\nu} \biggr)_{t=0} \label{ovrlpb}
\end{equation}
for some powers $n$. As is standard in pertuabtion theory in quantum
mechanics, if we use the wave functions of the unperturbed theory the
computation is only reliable for the first power $n$ for  which
(\ref{ovrlpb}) is non-zero. For all higher $n$'s one would need to
consider whether corrections to the wave functions themselves are 
relevant. Fortunately, as we will see shortly, the lowest nonzero
order of perturbation theory is precisely
$n=k-2$, and this is the term in which we are interested (as we can 
identify by quantum numbers).

We can simplify the expression considerably. First of all, it is easy
to see that the overlap is zero unless $g=g'h$, where $h$ is an
$SU(k-2)$ transformation. This is so because we start with a wave
function localized on the submanifold (\ref{adhmgauge}). If
$g\not=g'h$, then $g$ and $g'$ transport this function to two
different other submanifolds, and the overlap is zero. Hence in
(\ref{ovrlpb}) we can set $g=g'h$ and, using the $SU(k-2)$ invariance
of the functions, carry out only one integration
\begin{equation}
\int_g dg \, 
 \biggl(
{}_{\nu\bar\nu}\langle O_2| \times {}_{\eta\bar\eta}\langle O_2| \times f^*(Q)
 \biggl )_{t=0} \, || \, g^\dagger (\delta H + \delta H^{\dagger})^n g \, || \,
\biggl(
f(Q) \times {|O_2\rangle}_{\eta\bar\eta} \times {|O_k\rangle}_{\nu\bar\nu}
\biggr)_{t=0}
\label{ovrlpc}
\end{equation}

To simplify the computation further let us rewrite the state ${\vert
O_k \rangle}_{\nu\bar\nu}$ in a different way.   Each of the $N$ flavor
indices appears twice, hence we can write the state as
\begin{equation}
T_{A_1B_1..A_{k-2}B_{k-2}}
\nu_1^{A_1} \, \nu_1^{B_1} \cdots \nu_{k-2}^{A_{k-2}} \,
\nu_{k-2}^{B_{k-2}} \,
\vert \Phi \rangle \label{thestate1}
\end{equation}
for some tensor $T$.  This state is antisymmetric under the exchange of
each of the pairs $A_i\leftrightarrow B_i$ independently.  Hence
each pair either forms a ${\bf 5}$ or a singlet of $SO(5)$.  On the
other hand the total state is in the $(k-2)$-th symmetric traceless
rep of the $SO(5)$.  Therefore, each $A_i-B_i$ pair must be contracted
into a state in the ${\bf 5}$ of $SO(5)$. We have also explained
before (see the discussion around (\ref{okstat})) that each of the
$A_i$ and $B_i$ indices are either 1 or 2. Hence the state has to be
\begin{equation}
\nu_1^1 \, \nu_1^2 \cdots \nu_{k-2}^1 \, \nu_{k-2}^2 \,
\vert\Phi\rangle \, .
\label{thestate2}
\end{equation}

Having set up the states and the perturbation to the Hamiltonian, the
computation is very short. Let us first specify which power $n$ is
relevant in (\ref{ovrlpc}):

\begin{enumerate}
\item The interaction terms we wrote down do not involve $Q$,
hence the overlap $\langle f(Q)\vert f(Q)\rangle$ is straightforward,
and independent of the VEVs $\lambda$. It will contribute an irrelevant overall
factor.

\item Both in $|O_2 \rangle$ and $|O_k\rangle$ we chose the
same wavefunction in the ($\eta,\bar\eta$) Fock space. If we are
interested in the least number of insertions of $\delta H$, we can
take the direct overlap in this sector, i.e. ${}_{\eta\bar\eta}\langle
O_2| O_2 {\rangle}_{\eta\bar\eta}$.

\item Finally in the $\nu\bar\nu$ sector one proceed as
follows.  $g^\dagger (\delta H +\delta H^\dagger) g$ is a ${\bf 5}$ of
$SO(5)$. ${|O_2\rangle}_{\nu\bar\nu}$ is a singlet and
${|O_k\rangle}_{\nu\bar\nu}$ has charge $k-2$.  Therefore, in order to
get a non-zero 2-point function we need $k-2$ insertions of $g^\dagger
(\delta H +\delta H^\dagger) g$.  It is easy also to see that the
insertions should all be of $g^\dagger \delta H g$ and none of
$g^\dagger (\delta H)^\dagger g$.
\end{enumerate}

The crucial point is now that the insertion of $k-2$ powers of $\delta
H$ will give a result which is a rank $k-2$ holomorphic polynomial of
the entries of ${\lambda^{[12]}}^i_j$ (see (\ref{fininter})).  This
has precisely the correct quantum numbers to be the contribution of
the 1-point function of $O_{k-2}$ to the 2-point function that we are
computing!  What is more, it is the unique contribution of this
form.  It is reassuring that the only contribution that we can compute
reliably is indeed the one we are interested in, and that it is likely
to be protected by SUSY.

Hence the computation reduces to evaluation of the following matrix
element
\begin{equation}
\int_g {}_{\nu\bar\nu}\langle O_2 | \biggl( \Sigma_{i,j=3}^{k}
 (g^\dagger \lambda^{[12]} g)^i_j
(\nu^3_i {\bar\nu}^{4j} -\nu^4_i {\bar\nu}^{3j})\biggr)^{k-2}
{| O_k\rangle}_{\nu\bar\nu}
\label{finalintegral}
\end{equation}
Note that only the $(k-2)\times (k-2)$ lower right corner of the
matrix $g^\dagger\lambda g$, which we denote by $\Lambda_{(k-2)\times
(k-2)}$, participates in the interaction.  This is so because we need
to insert interaction terms which have only $\nu$ and $\bar\nu$, and
no $\eta$ and $\bar\eta$, and because the ($\nu,\bar\nu$) fermions are
the last $k-2$ entries of ($\mu,\bar\mu$) in (\ref{fininter}).

We would like to evaluate the integrand
\begin{equation}
    {}_{\nu\bar\nu}\langle O_2 | (\delta H)^{k-2} {| 
    O_k\rangle}_{\nu\bar\nu} =
{}_{\nu\bar\nu}\langle O_2 |
{\left(
\Sigma_{i,j} \Lambda^i_j (\nu^3_i {\bar\nu}^{4j}-\nu^4_i {\bar\nu}^{3j})
\right)^{k-2}}
{| O_k\rangle}_{\nu\bar\nu} 
\label{lstovlp}
\end{equation}
Let us first evaluate the function on a diagonal $\Lambda$.  Each
$\delta H$ is a sum over $\delta H_i=\Lambda^i_i (\nu^3_i
{\bar\nu}^{4i}-(4\leftrightarrow 3))$ for each flavor index $i$. 
Since each flavor index in $\vert O_k \rangle$ is  in
the ${\bf 5}$ of $SO(5)$, the only nonzero contribution from
$(\Sigma \delta H_i)^{k-2}$ is the term
\begin{equation}
N!\times \Pi_{i=1}^{k-2} \, \delta H_i \, .
\end{equation}
It is clear, therefore, that we get a result proportional to
\begin{equation}
\Pi_{i=1}^{k-2} \,  \Lambda^i_{i} \, , \label{result}
\end{equation}
 (It is also each to check that $\Pi_{i} \, \delta H_{i}$ does not
annihilate the states (\ref{thestate1}) so that the result is not
zero.)

Returning to a general, non-diagonal $\Lambda$, the expectation value
(\ref{lstovlp}) satisfies the following properties
\begin{enumerate}
\item It is a holomorphic polynomial of degree $k-2$.
\item It is $SU(k-2)$ invariant.
\item When $\Lambda$ is diagonal it is $N! \, \det (\Lambda)$.
\end{enumerate}
This determines it uniquely to be $N! \det(\Lambda)$ everywhere.

Reintroducing the integral over $g$ in (\ref{finalintegral}), and
rewriting $\det(\Lambda)$ in terms of the $\lambda$ matrix and $g$,
we get 
\begin{equation}
\int_g \, dg \, \det{}_{k-2} \,  (P \, g^\dagger \lambda^{[12]} g )
\label{almosttherea}
\end{equation}
where $P$ is the diagonal matrix $(0,0,1,1, \cdots1)$ and $\det{}_{k-2}$ 
is the subdeterminant introduced earlier.  The result of this integral is
proportional to
\begin{equation}
\det{}_{k-2} (\lambda^{[12]}) \, .
\label{there}
\end{equation}

\paragraph{Summary: } We computed the  expectation
value of the operator corresponding to a giant graviton along the
Coulomb branch of $\theory$ theory.  We chose the flat direction
$Z_i=\lambda_i$ and found that the expectation value of $O_{k-2}$
is the subdeterminant of the VEV $\det{}_{k-2}\lambda$.  This
supports our assertion that the giant gravitons should be
thought of as subdeterminants rather than traces.

\section{Discussion}
\label{discussion}

We have suggested a class of field theory operators, namely
subdeterminants, which corresponds to the giant gravitons in various
theories.  Many questions remain:
\begin{itemize}
    \item Is there a class of operators that roughly  extrapolates
    from traces to determinants, describing the graviton at
    each stage?  
   \item Can the scattering of a pointlike graviton from
    a giant graviton be computed in gravity?  Many such correlators
    are protected and should match the weak coupling computations that
    we have outlined.  
 \item There are also giant gravitons that
    expand on AdS rather than the sphere~\cite{myers,itsAki}.  These
    do not have a angular momentum bound.  How are they described as
    field theory states?  What is their relation to the giants on the
    sphere?  
\item Do determinants and subdeterminants play a role in the weak coupling
    `t Hooft planar expansion?
\end{itemize}
A classical gravitational background is in some sense a coherent state
of a large number of gravitons.  We might wonder whether such coherent
excitations of giant gravitons also exist.  Likewise, giants are
presumably associated with a generalization of reparametrization
invariance, in a sense the non-commutative extension of geometry in
the presence of higher rank gauge fields.  We have not addressed these
issues in this article, but understanding the dual representation of
giant gravitons is likely to be helpful.

\vspace{0.25in}
{\leftline {\bf Acknowledgements}}

We have enjoyed useful conversations and communications with Ofer
Aharony, Shmuel Elitzur, Rajesh Gopakumar, Esko Keski-Vakkuri, Igor
Klebanov, Albion Lawrence, Rob Myers, Ronen Plesser, Mukund Rangamani,
Simon Ross and Sandip Trivedi.  We thank Moshe Rozali for
collaboration on some of the material in Sec.~5.  {\small V.B.},
{\small M.S.} and {\small A.N.}  were supported by DOE grant
DOE-FG02-95ER40893.  {\small M.B.} was supported by the IRF Centers of
Excellence program, the US-Israel Binational Science Foundation, the
European RTN network HPRN-CT-2000-00122, the Minerva foundation
and the Einstein-Minerva center.
{\small V.B.} and {\small M.B.} are grateful to organizers of the
M-theory workshop at the ITP, Santa Barbara and the Amsterdam String
Theory Workshop for hospitality and free lunches.  {\small V.B.}
thanks the Weizmann Institute for hospitality while this work was
completed.

\appendix

\section{Expansion of the determinant in terms of traces}
\label{detexp}

The determinant of a matrix is  a product of eigenvalues, while the
trace is the sum.    Let $\phi$ be a matrix written in the eigenvector
basis so that it is diagonal. Then define
\[
u_k=Tr \phi^k
\]
Also, define $s_k$ to be the symmetric polynomials  in $a_i$ (where
$a_i$ are the entries in $\phi$ along the diagonal)
\[
s_k=(-)^k \sum_{i_1< \dots i_k} a_{i_1} \dots a_{i_k}
\]
Up to a sign, determinant of $\phi$ is $s_N$. The sets $u_k$ and $s_k$
are related by Newton's formula (see, e.g.,~\cite{detref})
\[
ks_k+\sum_{i=1}^k s_{k-i}u_i = 0,
\]
where $s_0=1$, $s_1=u_1=0$.
So we get the following system of equations:
\[
\left( \begin{array}{cccccccc}
N & 0 & u_2 & u_3 &  \cdots & u_{N-4}&u_{N-3} & u_{N-2} \\
0 & N-1 & 0 & u_2 &   \cdots & u_{N-5}&u_{N-4} & u_{N-3} \\
0 & 0 & N-2 & 0 & \cdots & u_{N-6}& u_{N-5} & u_{N-4} \\
. &.&.&.&.&.&.&.\\
.&.&.&.&.&.&.&.\\
0 & 0 & 0&0& \cdots  & 4 & 0 & u_2 \\
0 & 0& 0&0& \cdots  & 0 & 3 & 0 \\
0&0 & 0 &0& \cdots  & 0 & 0 & 2 \end{array} \right)
\left(\begin{array}{c} s_N \\ s_{N-1}\\  s_{N-2}\\ .\\.\\s_4\\s_3 \\s_2
\end{array} \right)= -\left(\begin{array}{c} u_N \\ u_{N-1}\\  u_{N-2}\\
.\\.\\u_4\\u_3 \\u_2
\end{array} \right)
\]

$N! s_N$ is the determinant of:\footnote{For a system
of equations
\[
Ax=b
\]
we know that the $jth$ components of $x=A^{-1}b$ is
\[
x_j=\frac{det \, B_j}{det \, A} \]
where $B_j$ is the same as matrix $A$ except that the vector $b$
replaces the $jth$ column of $A$. Notice that $ det \,A$ for our system
is $N!$}
\[
\left( \begin{array}{cccccccc}
-u_N & 0 & u_2 & u_3 &  \cdots & u_{N-4}&u_{N-3} & u_{N-2} \\
-u_{N-1} & N-1 & 0 & u_2 &   \cdots & u_{N-5}&u_{N-4} & u_{N-3} \\
-u_{N-2} & 0 & N-2 & 0 & \cdots & u_{N-6}& u_{N-5} & u_{N-4} \\
. &.&.&.&.&.&.&.\\
.&.&.&.&.&.&.&.\\
-u_4 & 0 & 0&0& \cdots  & 4 & 0 & u_2 \\
-u_3 & 0& 0&0& \cdots  & 0 & 3 & 0 \\
-u_2&0 & 0 &0& \cdots  & 0 & 0 & 2 \end{array} \right)
\]
The calculation of this determinant is not trivial.   We have computed
a few terms (the overall sign depends on
$N$):
\begin{equation}
det\,\phi = \frac{1}{N} u_N - \frac{1}{p(N-p)} u_p u_{N-p}
\dots -\frac{1}{N(N-k)(N-2k) \dots k} Tr (\phi^k)^{\frac{N}{k}}
\label{dettraceeq}
\end{equation}

\section{Bounds on the two point function of $O_{N/2} O_{N/2}$}
\label{bounds}

We wish to compute the two-point function $C_{2} = \langle (O_{N/2}
O_{N/2}) \, (O_{N/2}  O_{N/2})\rangle$:
\begin{eqnarray*}
 C_{2} & =& \frac{1}{((N/2)!)^4}
\epsilon_{i_1 \cdots i_{N/2}a_1 \cdots a_{N/2}}\,
\epsilon^{j_1 \cdots j_{N/2}a_1 \cdots a_{N/2}}\,
\epsilon_{l_1 \cdots l_{N/2}b_1 \cdots b_{N/2}}\,
\epsilon^{m_1 \cdots m_{N/2}b_1 \cdots b_{N/2}}\,\\&&
\epsilon_{p_1 \cdots p_{N/2}c_1 \cdots c_{N/2}}\,
\epsilon^{q_1 \cdots q_{N/2}c_1 \cdots c_{N/2}}\,
\epsilon_{r_1 \cdots r_{N/2}d_1 \cdots d_{N/2}}\,
\epsilon^{s_1 \cdots s_{N/2}d_1 \cdots d_{N/2}} \\&&
\langle:(\Phi^{i_1}_{j_1} \cdots \Phi^{i_{N/2}}_{j_{N/2}}
\Phi^{l_1}_{m_1} \cdots \Phi^{l_{N/2}}_{m_{N/2}})::
(\Phi^{p_1}_{q_1} \cdots \Phi^{p_{N/2}}_{q_{N/2}}
\Phi^{r_1}_{s_1} \cdots \Phi^{r_{N/2}}_{s_{N/2}}):\rangle \\
& > &\frac{1}{((N/2)!)^4}   \,\,\, 2(N/2)!^2 \\ & &
\epsilon_{i_1 \cdots i_{N/2}a_1 \cdots a_{N/2}}\,
\epsilon^{j_1 \cdots j_{N/2}a_1 \cdots a_{N/2}}\,
\epsilon_{l_1 \cdots l_{N/2}b_1 \cdots b_{N/2}}\,
\epsilon^{m_1 \cdots m_{N/2}b_1 \cdots b_{N/2}}\,\\&&
\epsilon^{i_1 \cdots i_{N/2}c_1 \cdots c_{N/2}}\,
\epsilon_{j_1 \cdots j_{N/2}c_1 \cdots c_{N/2}}\,
\epsilon^{l_1 \cdots l_{N/2}d_1 \cdots d_{N/2}}\,
\epsilon_{m_1 \cdots m_{N/2}d_1 \cdots d_{N/2}}
\\
& = & \frac{1}{((N/2)!)^4}   \,\,\, 2(N/2)!^2 \,\,\,(N/2)!^4\,\,N!^2
\end{eqnarray*}
There is a $>$ sign because the r.h.s. is the result of only
a subset of the possible contractions. Specifically, we are looking
at only those contractions in which all the $\Phi^i_j$'s are contracted
with all the $\Phi^p_q$'s etc., and are ignoring the contractions of the type
in which some of the $\Phi^i_j$ is contracted with $\Phi^p_q$ and some with
$\Phi^r_s$. The factor $\frac{1}{((N/2)!)^4} $ comes from the definition
of $O_{N/2}$,  $ 2(N/2)!^2$ is the number of possible contractions of
the particular type we are keeping and $(N/2)!^4\,\,N!^2$ is the result
of summing over the indices of the eight epsilon tensors.

The upper bound is obtained by assuming that all of the $N!$ possible
contractions give the same result after the espilons are contracted.
The result of summing the eight possible epsilons is bounded above
by $(N/2)!^4\,\,N!^2$ so the upper bound on our correlator is
\[
 \frac{1}{((N/2)!)^4}   \,\,\, (N!) \,\,\,(N/2)!^4\,\,N!^2=N!^3
\]


\end{document}